\documentclass[reprint,amsmath,amssymb,aps,floatfix]{revtex4-1}
\usepackage{graphicx}
\usepackage{dcolumn}
\usepackage{bm}
\usepackage{amsmath}
\usepackage{amssymb}
\usepackage{txfonts}
\usepackage{url}
\usepackage{comment}
\usepackage{tabularx}
\usepackage{multirow}
\usepackage[toc,page]{appendix}
\usepackage[usenames,dvipsnames]{color}
\definecolor{Gray}{gray}{0.0}
\definecolor{lightGray}{gray}{0.35}

\newcommand{\etal}{\textit{et al.}}

\begin{document}

\setlength{\abovedisplayskip}{0pt}
\setlength{\belowdisplayskip}{0pt}
\setlength{\abovedisplayshortskip}{0pt}
\setlength{\belowdisplayshortskip}{0pt}

\title{\emph{Ab initio}
  thermodynamic properties of certain compounds in Nd-Fe-B system}
\author{Adie Tri Hanindriyo$^{1,*}$} \author{Soumya Sridar$^{2}$}
\author{K.C. Hari Kumar$^{2}$} \author{Kenta Hongo$^{3,4,5,6}$} \author{Ryo
  Maezono$^{3,7}$}

\affiliation{}

\affiliation{$^{1}$ School of Material Science,
  JAIST, Asahidai 1-1, Nomi, Ishikawa, 923-1292, Japan }

\affiliation{$^{2}$ Department of Metallurgical and Materials Engineering,
  Indian Institute of Technology Madras, Chennai-600 036, India }

\affiliation{$^{3}$ Computational Engineering Applications Unit, RIKEN, 2-1
  Hirosawa, Wako, Saitama, 351-0198, Japan }

\affiliation{$^{4}$ Research Center for Advanced Computing Infrastructure,
  JAIST, Asahidai 1-1, Nomi, Ishikawa, 923-1292, Japan }

\affiliation{$^{5}$ Center for Materials Research by Information
  Integration, Research and Services Division of Materials Data and
  Integrated System, National Institute for Materials Science, Tsukuba
  305-0047, Japan }

\affiliation{$^{6}$ PRESTO, Japan Science and Technology Agency, 4-1-8
  Honcho, Kawaguchi-shi, Saitama 332-0012, Japan }

\affiliation{$^{7}$ School of Information Science, JAIST, Asahidai 1-1,
  Nomi, Ishikawa, 923-1292, Japan}

\affiliation{}

\affiliation{$^{*}$ adietri@icloud.com }

\begin{abstract}
  In this work, we report the results of \emph{ab initio} calculations of
  thermochemical properties of several compounds in the Fe-Nd, B-Nd and
  B-Fe-Nd systems. We have performed DFT+U calculations to compute the
  enthalpy of formation of the compounds NdB$_6$, NdB$_4$, Nd$_2$B$_5$,
  Nd$_2$Fe$_{17}$ and Nd$_5$Fe$_2$B$_6$.  It was found that the values
  obtained with an effective Hubbard $U$ correction have better agreement
  with the experimental data. We have also computed the vibrational
  contribution to the heat capacity ($C_p$) of the compounds as a function
  of temperature was computed using the quasiharmonic approximation. For
  most of the compounds these properties have not been experimentally
  determined until now. Electronic contribution to the heat capacity has
  been calculated as well from electronic structure calculations.
  A part of the results in this work has been applied to the re-optimization
  of the binary Nd-B phase diagram.
\end{abstract}
\maketitle

\section{Introduction}
\label{sec.intro}
\emph{Ab initio} calculations based on density functional theory (DFT) have
gained much popularity mainly because of the reliable results that can
be obtained through these computations.  Amongst many properties that can
be computed by this method, the thermochemical properties such as enthalpy
of formation, enthalpy of mixing, heat capacity, \emph{etc.} are of
particular interest. They serve as important input in the Gibbs energy
modelling in the context of the CALPHAD (acronym for CALculation of PHAse
Diagrams) approach, especially when the corresponding experimental data are
not available. This greatly improves the quality of the generated Gibbs
energy functions. With the advancements in high-performance computing, it
is now possible to carry out these calculations for systems with a large
number of atoms. In this work, we report the results of \emph{ab initio}
calculations of thermochemical properties of the compounds NdB$_6$,
NdB$_4$, Nd$_2$B$_5$, Nd$_2$Fe$_{17}$ and Nd$_5$Fe$_2$B$_6$.

The rare-earth and transition metal compounds with localized $3d/4f$
orbitals constitute what is known as the ``strongly correlated''
systems. Compounds of this class contain localized electronic orbitals that
introduce significant on-site Coulomb repulsion.~\cite{1991ANI} Mott
insulators are also a well-known group of materials belonging to this
category.

A shortcoming of DFT is the failure of the conventional
exchange-correlation functionals (LDA/GGA) in describing the quantum
mechanical nature of the strongly correlated systems.  For instance, the
DFT calculations using LDA/GGA functionals identify Mott insulators
as metals. On the other hand, Mott insulators are well described by the
Hubbard or Anderson models based on the tight-binding
approach.~\cite{1991ANI} The $U$ parameter in the Hubbard model accounts
for the effect of on-site repulsion. This parameter is adopted within DFT
and gives rise to the so-called DFT+U method, which introduces a correction
to the total energy.~\cite{1997ANI} This approach has thus increased the
accuracy of the results of DFT calculations concerning strongly correlated
materials. These orbitals are specially treated in the DFT+U scheme with
the correction using a Hubbard $U$ parameter.

To the best of our knowledge, \emph{ab initio} calculations to compute the
finite temperature thermodynamic properties of these compounds have not
been reported so far. Li \etal{}~\cite{2014LI} have reported \emph{ab
  initio} thermodynamic properties of several borides relevant to the Fe-B
system.  Recently, another \emph{ab initio} investigation~\cite{2018COL}
reported enthalpies of formation of Nd-B compounds using GGA functionals.

The thermodynamic modelling of Nd-Fe-B system by van Ende \etal{}
~\cite{Ende:2013} made use of the available experimental $C_p$ data for
Gibbs energy modelling of NdB$_6$ and NdB$_4$, whereas $C_p$ of
Nd$_2$B$_5$, Nd$_2$Fe$_{17}$ and Nd$_5$Fe$_2$B$_6$ were approximated using
Neumann-Kopp rule due to lack of experimental data. It is envisaged that
the results of the present work could be used in order to further improve
the thermodynamic modelling of Nd-Fe-B system.

\section{Model and Methodology}
\label{sec.det.spec}
The ordered compounds studied in this work are parts of the stable phases classically defined
within the Nd-Fe-B ternary system.
Existence of several ordered compounds are conclusively established within previous works,
such as the 4 compounds in the binary Nd-B system (NdB$_6$, NdB$_4$,
Nd$_2$B$_5$, and NdB$_{66}$), while others, such as Nd$_5$Fe$_{17}$, were more recently
discovered.\cite{1990MOR}
Within this work, we have included ordered phases listed in the latest assessment of the Nd-Fe-B
system in 2013 by Ende and Jung~\cite{Ende:2013}, with exception of the thoroughly investigated
Fe-B system.
This includes 4 phases in the binary Nd-B system, 2 phases in the Nd-Fe system, and 3 established
phases in the Nd-Fe-B ternary system.

\vspace{2mm}
Several among these phases possess incommensurate crystal structures, which unfortunately
necessitates a large unit cell and is impractical for first-principles calculation with the
available computational resources.
For example, the aforementioned Nd$_5$Fe$_{17}$ phase possess an incommensurate crystal structure
which requires a quite large primitive cell (\~150 atoms per unit cell).
Computational cost limitations play a key part in determining the scope of this study,
with such compounds and others with similarly complicated or long-range ordered crystal
structures (NdB$_{66}$, Nd$_{1.1}$Fe$_4$B$_4$) excluded from our investigation.
This is a key factor in this study as we are aiming for not only the formation enthalpy,
but also the finite temperature heat capacity by phonon calculations, which already
requires significant computational cost to begin with.
Metastable compounds such as the ternary Nd$_2$Fe$_{23}$B$_3$~\cite{1993GOU} are also excluded due to both
computational cost and phase stability reasons (the trend to ignore such compounds
in the latest assessments).
As a result, we settled on three binary Nd-B compounds (NdB$_6$, NdB$_4$, and Nd$_2$B$_5$),
one binary Nd-Fe compound (Nd$_2$Fe$_{17}$), and another ternary Nd-Fe-B compound (Nd$_5$Fe$_2$B$_6$,
referred to as T3 in the work of Ende and Jung~\cite{Ende:2013}).
These compounds are suitable for our available computational facilities available at JAIST.

\vspace{2mm}
In this work, DFT calculations were carried out using the QUANTUM
ESPRESSO~\cite{Gianozzi:2009} code, which uses a plane-wave basis set to
expand the wavefunctions. Core electrons were represented with projector
augmented waves (PAW)~\cite{Blochl:1994} pseudopotentials. These
pseudopotentials use 3 valence electrons (2s$^2$ 2p$^1$) for B, 8 valence
electrons (4s$^2$ 3d$^6$ 4p$^0$) for Fe, and 14 valence electrons (5s$^2$
5p$^6$ 6s$^2$ 4f$^4$ 5d$^0$) for Nd. The GGA exchange-correlation
functional parametrized by Perdew-Burke-Ernzerhof~\cite{Perdew:1996} (PBE)
was used. Marzari-Vanderbilt smearing~\cite{Marzari:1999} was used to
smooth the discontinuities in integrals due to the Fermi surface. To aid
convergence in calculations, the largest smearing width $\sigma$ was used
which does not introduce ground state energy deviation larger than 0.002
Ry/atom. The width of Marzari-Vanderbilt smearing used in this study ranged
from 0.01 Ry up to 0.05 Ry.

\vspace{2mm}
The calculations were done on $k$-points selected using the Monkhorst-Pack
scheme.~\cite{Monkhorst:1976} Convergence tests were carried out for every
structure up to an energy convergence of 0.001 Ry/atom. The same
convergence test was applied to determine the appropriate cut-off energy
for the plane-wave expansion of the wavefunction. The Hubbard correction
(GGA+U) was employed to account for localized orbitals due to the presence
of $3d/4f$ orbitals.  The effective $U$ corrections are obtained using an
effective $U$ parameter $U_{\rm{eff}}=U-J$ as outlined by Dudarev
\etal{}.~\cite{1998DUD}

\vspace{2mm}
The Hubbard $U$ parameter allows for the self-consistent determination as
outlined and implemented by Cococcioni
\etal{}.~\cite{Cococcioni:2005} Since DFT treats orbitals as delocalized,
partial occupation is favoured. Hubbard $U$ correction can be considered as
the energy cost to enforce full occupation of the Hubbard sites. Cococcioni
\etal{} formulated a linear response approach by perturbation of the
occupation matrices for the Hubbard sites. By introducing a finite
perturbation in the occupation matrix, the gradient of energy with respect
to orbital occupation may be established. Hence, the energy cost of full
occupation may be obtained, which is the effective Hubbard $U$ parameter.

\vspace{2mm}
Calculations were performed using GGA+U with self-consistently determined
values for $U_{\rm{eff}}$. These were obtained for every inequivalent
atomic sites occupied by the Nd and Fe. The Nd or Fe atoms occupying each
inequivalent site were perturbed separately and $U_{\rm eff}$ obtained for
the corresponding Hubbard site. The magnitude of $U_{\rm eff}$ is related
to the atom-like nature of electrons occupying the Hubbard sites.

\vspace{2mm}
The enthalpy of formation for different compounds were calculated with
respect to ($\alpha$-Nd), ($\alpha$-Fe) and ($\alpha$-B) using the following
equation.

\begin{align}\label{Hform1}
  \Delta _{\text{f}}H_{298}^\circ
  ({\text{Nd}_x}{\text{Fe}_y}{{\text{B}}_z}) \approx\ &\ {E_0}({\text{Nd}_x}{\text{Fe}_y}{{\text{B}}_z})\nonumber\\
   & - x{E_0}({\text{Nd}}) - y{E_0}({\text{Fe}}) - z{E_0}({\text{B}})
\end{align}
\\
In order to compute the vibrational contribution to $C_p$ of the compounds,
phonon frequencies were computed using the quasiharmonic approximation
(QHA), as implemented in the phonopy~\cite{2015TOG} code. The approach uses
finite displacements in order to compute force constants, which are
subsequently used to construct the dynamical matrices and phonon
frequencies. Vienna Ab Initio Simulation Package
(VASP)~\cite{Kresse:1993,Kresse:1994,Kresse:1996_1,Kresse:1996_2} was used
as the force calculator, since it provides more reliable geometry
optimization for QHA purposes. Phonon calculations are performed for these
compounds using varying supercell sizes. Choices regarding computed supercell
sizes are made with respect to computational cost and the number of calculations
needed over the period of study.

\vspace{2mm}
There are two formulations of heat capacity ($C_p$) as implemented in
phonopy.~\cite{2010TOG} The phonon contribution to Helmholtz free energy is
computed as in Equation~\eqref{Ffree}:

\begin{align}\label{Ffree}
  F_{\rm{phonon}} = &\frac{1}{2}\sum\limits_{\bf{q}\nu}\hbar\omega({\bf{q}}\nu)\nonumber\\
  &+k_BT\sum\limits_{\bf{q}\nu}\ln\left[1-\exp\left(-\frac{\hbar\omega({\bf{q}}\nu)}{k_BT}\right)\right]
\end{align}
\\
where $\nu$ denotes band index and ${\bf{q}}$ denotes the q-points.
The Gibbs energy is used in order to derive properties at constant pressure.
It may be calculated as the minimum of cell volume $V$, as seen in Equation~\eqref{Gfree}:

\begin{align}\label{Gfree}
  G\left(T,p\right) = \min_{V}\left[U(V)+F_{\rm{phonon}}(T;V)+pV\right]
\end{align}
\\
which requires the computation of the Helmholtz free energy at multiple volumes.
This is applied over a volume range of $-5\%$ to $+5\%$ of the
equilibrium volume, with 1\% increments (a total of 11 points).
$C_p$ can then be calculated from the second derivative of Gibss free energy
with respect to temperature:
\begin{align}\label{Cp}
  C_p\left(T,p\right) = -T\frac{\delta^2 G\left(T,p\right)}{\delta T^2}
\end{align}
\\
The result of Equation~\eqref{Cp} is collated in the file Cp-temperature.dat within
phonopy code. A different derivation may produce a smoother result in the
Cp-temperature\_polyfit.dat file within phonopy, as described in Equation~\eqref{polyfit}:

\begin{align}\label{polyfit}
  C_p\left(T,p\right) = &\left.T\frac{\delta V\left(T,p\right)}{\delta T}\frac{\delta S\left(T;V\right)}{\delta V}\right|_{V(T,p)}\nonumber\\
  &+C_V\left[T,V(T,p)\right]
\end{align}
\\
where $S$ denotes the entropy, again taken at minimum with respect to volume $V$ at
temperature $T$.

\vspace{2mm}
While the vibrational contribution to heat capacity is the most significant factor,
inclusion of the electronic contribution to heat capacity is also important for metallic compounds,
especially for the high $T$ region.
The electronic contribution for metallic compounds is often approximated from the
electron gas model:

\begin{align}\label{elec}
  C_{\rm elec}\left(T\right) = \frac{\pi^2}{3}k_{B}^2TD(E_{F})
\end{align}
\\
where $D(E_F)$ is the density of states at the Fermi level.
The density of states are calculated by so-called non-SCF (non-self consistent field)
calculations making use of the ground state charge density obtained from the previous
single point or SCF calculations, with a much finer k-points grid.
The k-point grids used in these calculations are set at 3 times the density
of the SCF calculations.
It is common practice among CALPHAD researchers to develop thermodynamic functions valid
up to $T = 3000$ K for a complete database.

\vspace{2mm}
Heat capacity obtained by phonon calculation reflects only the contribution of lattice (vibrational)
effects, which accounts for most of the heat capacity of materials.
Electronic contribution, meanwhile, is significant for metallic compounds in the high $T$ region.
Other contributions, such as magnetic contributions, are considered relatively insignificant
for the compounds in this study.
This is due to the mathematical nature of each contribution, especially pertaining to the
temperature ranges considered for CALPHAD purposes.
The thermodynamic functions for unaries within CALPHAD are valid only above room temperature,
which necessarily limits the investigation to values of $T \geq 300$ K.
Lattice contributions scale with temperature by a cubic $T^3$ term, which explains how it accounts
for most of the heat capacity up to the Debye temperature, and the electronic contributions scale linearly
with $T$ for metallic systems which makes it significant for higher temperatures, while magnetic
contributions, responsible for a $T^{3/2}$ term, play a key part in the low temperature behavior of
$C_p$.~\cite{2013VAR}
This behavior lends credence to the importance of calculation of lattice vibrations and electronic
contributions in this work for CALPHAD purposes, for the region of $T \geq 300$ K.

\vspace{2mm}
In first principles calculations, materials are treated at their respective ground states.
This includes the magnetic ordering at 0 K, which is reflected in the energetics: the ground
state magnetic ordering will produce the lowest total energy in the calculation.
This also means that any changes in the magnetic ordering at finite temperatures (as is the norm
for magnetic materials) is not reflected in the phonon calculation results.
The most striking example of this phenomenon is the small peaks in $C_p$ that is often seen in materials
which undergo changes in magnetic ordering at some finite temperature (Neel, Curie, etc.),
which is an often seen phenomenon in magnetic materials.
These peaks, which signify different magnetic phases, are most commonly observed in low $T$ regions
where contributions from magnetic moments are most significant.
They are completely missed, therefore, in heat capacity predictions which take into account only
lattice vibration and electronic contributions.

\vspace{2mm}
The choice of magnetic ordering at ground state in phonon calculation serves to more accurately
determine the optimal geometry, which is essential in the harmonic approximation.
Within the harmonic approximation~\cite{2015TOG}, before any finite perturbations are introduced,
the resultant forces within quantum systems are approximated as zero.
Therefore, it is vital to perform the best possible geometry optimization and drive the
interatomic forces closer to zero as possible.
Accurate magnetic ordering at 0 K serves to better reflect the interatomic interactions
within the crystal structure and to obtain better optimized structures.
In this work, we have made efforts to model ground states of Nd-Fe-B compounds which best
reflect the 0 K magnetic ordering discovered in previous experimental results.
This is also the key to successful computations since certain magnetic orderings may produce
unstable structures at 0 K, which will appear in the phonon dispersion as negative
frequencies, potentially leading to flawed prediction of the lattice contribution.

\vspace{2mm}
Both bulk Nd and NdB$_6$ were modeled with antiferromagnetic ordering at 0 K, along
the [111] plane for cubic NdB$_6$.
NdB$_4$, meanwhile, possesses complex multipolar ordering which undergoes several
transitions at low $T$ and an antiferromagnetic quadrupolar (AFQ) ground
state.~\cite{Yamauchi:2017}
It is approximated using a simple antiferromagnetic ordering along the [110] plane.
Meanwhile, Nd$_2$Fe$_{17}$ possesses finite magnetic moment~\cite{Long:1994} according to
measurement at 10 K, and as such is constrained to the ferromagnetic ordering.
Limited information is available for some phases, unfortunately.
The atomic magnetic momenta for these compounds are allowed to freely optimize during both
self-consistent calculations and geometry optimization, unconstrained to any one specific
ordering.
This is true for two compounds, Nd$_2$B$_5$ and Nd$_5$Fe$_2$B$_6$, in this work.

\vspace{2mm}
Magnetic contribution scales with $T^{3/2}$ and provides significant contribution
at low $T$ and at magnetic phase transitions.
Apart from these cases, a phonon contribution far outweighs both magnetic and
electronic contributions up to the Debye temperature, and is expected to dominate for
temperature regime above 300 K.
Therefore, the results of phonon calculation in this work is predicted to be
most relevant for the temperature range in question.

\vspace{2mm}
For visualization of crystal structures, the XCRYSDEN code~\cite{1999KOK} was
used. The initial crystal structures used in all our calculations are well defined
in Tables in Section~\ref{sec.results}, while the results of our geometric optimization
calculations (performed for every compound), as well as the computational parameters
involved, are listed in the Appendix.

\section{{\emph{Ab initio}} Calculation Results and discussion}
\label{sec.results}
\subsection{Elements}
\subsubsection{Nd}
The ground state crystal structure of Nd ($\alpha$-Nd) has the $P6_3/mmc$
symmetry. Nd atoms are arranged in a double hexagonal close-packed
(dhcp) arrangement with two distinct symmetry-irreducible positions with an
antiferromagnetic ordering.
In order to account for the antiferromagnetic ordering in the structure,
the initial magnetic moments for the two positions are specified as antiparallel
to one another.

\begin{figure*}
  \begin{center}
    \includegraphics[width=250pt]{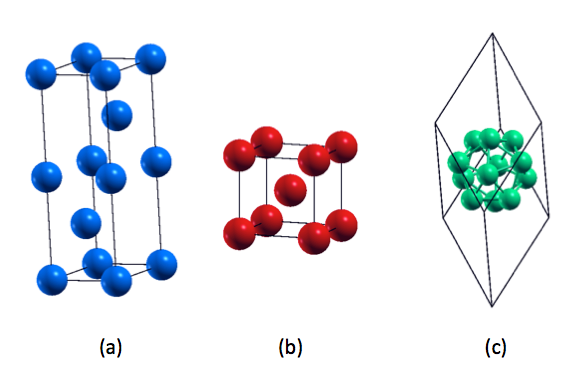}
    \caption{Crystal structure of (a) $\alpha$-Nd, (b) $\alpha$-Fe and (c)
      $\alpha$-B visualized using XCRYSDEN~\cite{1999KOK}}
    \label{fig:Elements}
  \end{center}
\end{figure*}

\vspace{2mm}
The crystal structure is shown in Figure~\ref{fig:Elements}.
Initial data on the lattice parameters and atomic positions are taken from the work of
Wyckoff~\cite{1963WYC} (Table~\ref{tab:Nd_str}). The calculated $U_{\rm{eff}}$ values per
the scheme of Cococcioni and de Gironcoli~\cite{Cococcioni:2005} are also included.

\begin{table}
  \begin{center}
    \caption{Crystallographic data for $\alpha$-Nd}
    \begin{tabular}{l l c c c c}
      \hline
      Space group && \multicolumn{4}{l}{$P6_3/mmc$ (194)}\\
      Pearson symbol && \multicolumn{4}{l}{$hP4$}\\
      $a_0 [\AA]$ && \multicolumn{4}{l}{3.6582}\\
      $c_0 [\AA]$ && \multicolumn{4}{l}{11.7966}\\
      \hline
      \multirow{2}{*}{Label} & \multirow{2}{*}{Type} & \multicolumn{3}{c}{Fractional Coordinates} & \multirow{2}{*}{$U_{\rm{eff}}$ [eV]}\\
                  & & \phantom{0000}$x$\phantom{0000} & \phantom{0000}$y$\phantom{0000} & \phantom{0000}$z$\phantom{0000} &\\
      \hline
      Nd1 & Nd & 0.00000 & 0.00000 & 0.00000 & 5.1340\\
      Nd2 & Nd & 0.33333 & 0.66667 & 0.25000 & 5.2474\\
      \hline
    \end{tabular}
    \label{tab:Nd_str}
  \end{center}
\end{table}

\subsubsection{Fe}
The ground state crystal structure of Fe ($\alpha$-Fe) belongs to the
$Im\bar{3}m$ space group. The crystal structure is shown in
Figure~\ref{fig:Elements}, with a ferromagnetic ordering. The initial
lattice constants were taken from Kohlhaas \etal{}.~\cite{Kohlhaas:1967}
The details of the crystal structure is listed in Table~\ref{tab:Fe_str}.
A single Hubbard site was treated with $U_{\rm{eff}}$ of~3.9182 eV.

\vspace{2mm}
From geometry optimization using GGA, we obtained an underestimated lattice
constant of around 2.8253~$\AA$. This result is comparable to recent
\emph{ab initio} calculations of ground state Fe, which shows a slight
underestimation.~\cite{2011JAN,2010KOD} However, from geometry optimization using GGA+$U$ an
overestimated lattice constant of around 2.9127~$\AA$ was obtained, leading
to a large difference in ground state energy between the two optimized structures.
This would significantly affect enthalpy of formation calculations for both Nd$_2$Fe$_{17}$ and
Nd$_5$Fe$_2$B$_6$, discussed in Sections~\ref{sec:Nd2Fe17} and~\ref{sec:Nd5Fe2B6}.

\begin{table}
  \caption{Crystallographic data for $\alpha$-Fe}
  \begin{center}
    \begin{tabular}{l l c c c c}
      \hline
      Space group && \multicolumn{4}{l}{$Im\bar{3}m$ (229)}\\
      Pearson symbol && \multicolumn{4}{l}{$cI2$}\\
      $a_0 [\AA]$ && \multicolumn{4}{l}{2.8665}\\
      \hline
      \multirow{2}{*}{Label} & \multirow{2}{*}{Type} & \multicolumn{3}{c}{Fractional Coordinates} & \multirow{2}{*}{$U_{\rm{eff}}$ [eV]}\\
                  & & \phantom{0000}$x$\phantom{0000} & \phantom{0000}$y$\phantom{0000} & \phantom{0000}$z$\phantom{0000} &\\
      \hline
      Fe1 & Fe & 0.00000 & 0.00000 & 0.00000 & 3.9182\\
      \hline
    \end{tabular}
    \label{tab:Fe_str}
  \end{center}
\end{table}

\subsubsection{B}
There is uncertainty regarding the ground state structure of boron, whether
it is $\alpha$-B or $\beta$-B.  Both $\alpha$-B and $\beta$-B belong to the
space group $R\bar{3}m$. While $\alpha$-B was found to be energetically more
stable in an earlier work by Shang \etal{}~\cite{Shang:2007}, another study
found $\beta$-B to be more stable by an energy difference of
3~meV/atom.~\cite{Setten:2007} The $\alpha$-B is regarded as the ground
state in this work due to this marginal difference in energy and its
much simpler structure compared to $\beta$-B. The initial lattice parameters and
atomic positions required for the calculations are taken from the
experimental work of Will and Kiefer.~\cite{Will:2001} These are listed in
Table~\ref{tab:B_str}. A rhombohedral primitive cell was constructed based
on the crystal structure data found in their work. It contains 12 atoms
(Figure~\ref{fig:Elements}), compared to the 105 atoms in the case of
$\beta$-B.

\begin{table}
  \caption{Crystallographic data for $\alpha$-B}
  \begin{center}
    \begin{tabular}{l l c c c c}
      \hline
      Space group && \multicolumn{4}{l}{$R\bar{3}m$ (166)}\\
      Pearson symbol && \multicolumn{4}{l}{$hR12$}\\
      $a_0 [\AA]$ && \multicolumn{4}{l}{4.9179}\\
      $c_0 [\AA]$ && \multicolumn{4}{l}{12.5805}\\
      \hline
      \multirow{2}{*}{Label} & \multirow{2}{*}{Type} & \multicolumn{3}{c}{Fractional Coordinates} & \multirow{2}{*}{$U_{\rm{eff}}$ [eV]}\\
                  & & \phantom{0000}$x$\phantom{0000} & \phantom{0000}$y$\phantom{0000} & \phantom{0000}$z$\phantom{0000} &\\
      \hline
      B1 & B & 0.45219 & 0.54781 & 0.05800 & -\\
      B2 & B & 0.53019 & 0.46981 & 0.19099 & -\\
      \hline
    \end{tabular}
    \label{tab:B_str}
  \end{center}
\end{table}

\begin{figure*}
  \begin{center}
    \includegraphics[width=250pt]{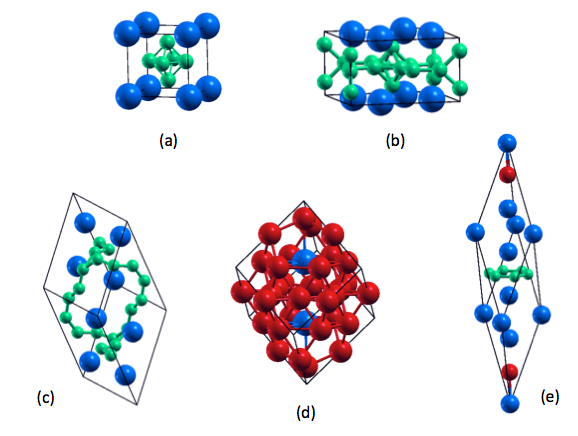}
    \caption{Primitive cells of (a) NdB$_6$, (b) NdB$_4$, (c) Nd$_2$B$_5$,
      (d) Nd$_2$Fe$_{17}$, and (e) Nd$_5$Fe$_2$B$_6$, visualized using
      XCRYSDEN. ~\cite{1999KOK}}
    \label{fig:Compounds}
  \end{center}
\end{figure*}

\subsection{NdB$_6$}
NdB$_6$ possesses with a cubic crystal structure with $Pm\bar{3}m$ symmetry
(Figure~\ref{fig:Compounds}). The initial lattice parameters
are taken from the work of McCarthy and Tompson~\cite{McCarthy:1980},
while the atomic positions are from the prototype structure CaB$_6$.~\cite{Han:2013} The
crystal structure data is listed in Table~\ref{tab:NdB6_str}. No
significant change in lattice parameters or atomic positions was observed
after the geometry optimization.

\begin{table}
  \caption{Crystallographic data for NdB$_6$}
  \begin{center}
    \begin{tabular}{l l c c c c}
      \hline
      Space group && \multicolumn{4}{l}{$Pm\bar{3}m$ (221)}\\
      Pearson symbol && \multicolumn{4}{l}{$cP7$}\\
      $a_0 [\AA]$ && \multicolumn{4}{l}{4.1280}\\
      \hline
      \multirow{2}{*}{Label} & \multirow{2}{*}{Type} & \multicolumn{3}{c}{Fractional Coordinates} & \multirow{2}{*}{$U_{\rm{eff}}$ [eV]}\\
                  & & \phantom{0000}$x$\phantom{0000} & \phantom{0000}$y$\phantom{0000} & \phantom{0000}$z$\phantom{0000} &\\
      \hline
      Nd1 & Nd & 0.00000 & 0.00000 & 0.00000 & 5.3377\\
      B2  & B  & 0.20170 & 0.50000 & 0.00000 & -\\
      \hline
    \end{tabular}
    \label{tab:NdB6_str}
  \end{center}
\end{table}

\vspace{2mm}
As a part of our initial calculations, the enthalpy of formation of NdB$_6$
was calculated using several methods. One of the calculations was done with
GGA-PBE exchange-correlation functional without any correction. The other
calculations involve Hubbard $U$ corrections and with inclusion of the $J$
term (GGA$+U+J$). The last calculation was performed with the effective $U$
parameter (GGA$+U_{\rm{eff}}$). This approach was done in order to
establish the reliability of the effective Hubbard correction, in
comparison to the non-corrected GGA and stricter Hubbard corrections.

\vspace{2mm}
GGA$+U+J$ calculation was performed with uniform correction
parameters. The $U$ value was taken as 4.8 eV and the $J$ value as 0.6 eV,
both taken from a cRPA (constrained random phase approximation) calculation
by Nilsson \etal{}.~\cite{Nilsson:2013} The effective Hubbard parameter
$U_{\rm{eff}}$ was estimated to be 5.337 eV using the method proposed by
Cococcioni.~\cite{Cococcioni:2005} The calculated enthalpy of formation
using the different methods are listed in Table~\ref{tab:NdB6}, along with the
experimental data from the work of Storms.~\cite{Storms:1981}

\begin{table}
  \caption{Comparison of theoretical and experimental enthalpy of formation
    for NdB$_6$}
  \begin{center}
    \begin{tabular}{l c}
      \hline
      Method & $\Delta _{\text{f}}H_{298}^\circ$ [J/mol] \\
      \hline
      GGA & -51720\phantom{ $\pm$ 1250}\\
      GGA$+U+J$\phantom{000} & -46650\phantom{ $\pm$ 1250}\\
      GGA$+U_{\rm{eff}}$ & -45019\phantom{ $\pm$ 1250}\\
      Experiment & -46750 $\pm$ 1250\\
      \hline
    \end{tabular}
    \label{tab:NdB6}
  \end{center}
\end{table}

\vspace{2mm}
The enthalpy of formation calculated using GGA$+U+J$ and GGA$+U_{\rm{eff}}$
methods are in good agreement with the experimental data. GGA-PBE
calculations show slightly more negative value. These results show that
while conventional GGA is sufficient to predict the enthalpy of formation
of NdB$_6$ to a certain extent, both the simplified and strict Hubbard
corrections show at least some measure of improvement. The Hubbard correction
allows for a better representation of the localized orbitals on the Hubbard
site (atomic-like $4f$ orbitals), leading to a more reliable prediction of
energetics of the system.

\begin{figure}
  \begin{center}
    \includegraphics[width=225pt]{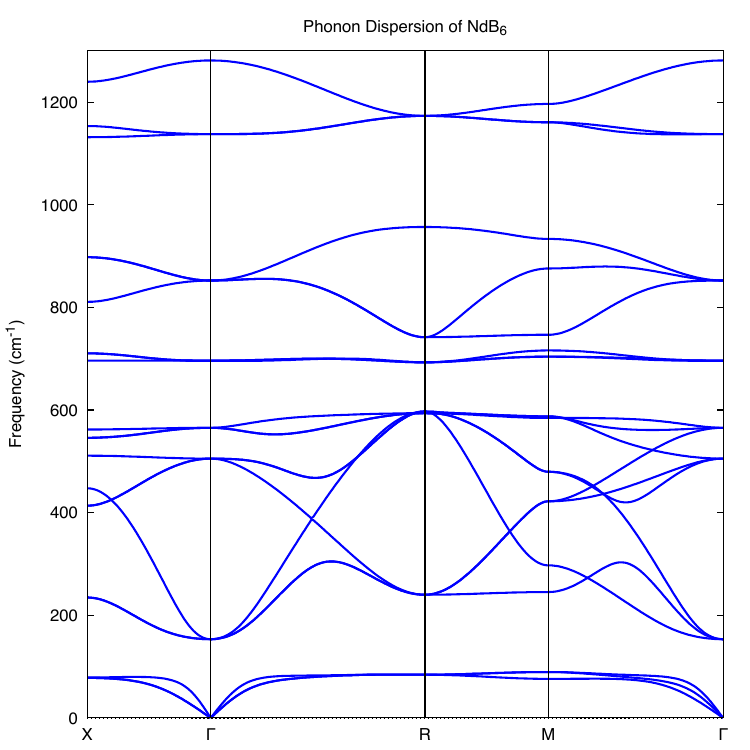}
    \caption{Calculated phonon dispersion for NdB$_6$.}
    \label{fig:NdB6_phon}
  \end{center}
\end{figure}

\begin{figure}
  \begin{center}
    \includegraphics[width=225pt]{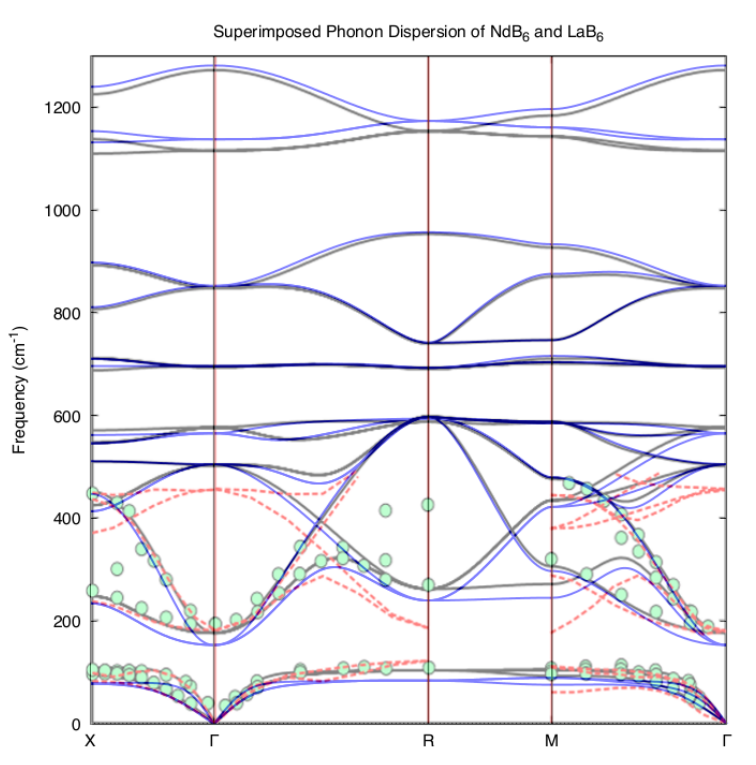}
    \caption{Comparison between calculated phonon dispersion of $1~\times~1~\times~1$ supercell NdB$_6$
      (blue) in this work and LaB$_6$. Three data sets for LaB$_6$ are
      shown, including DFPT calculation results by G{\"{u}}rel and
      Eryi${\breve{\rm{g}}}$it~\cite{Gurel:2010} (black), frozen phonon
      calculation results of Monnier and Delley~\cite{Monnier:2004} (dashed
      red), and the experimental work of Smith \etal{}~\cite{Smith:1985}
      (circles). Modified from Figure 2 of G{\"{u}}rel and
      Eryi${\breve{\rm{g}}}$it's work~\cite{Gurel:2010}}
    \label{fig:LaB6}
  \end{center}
\end{figure}

\begin{figure}
  \begin{center}
    \includegraphics[width=225pt]{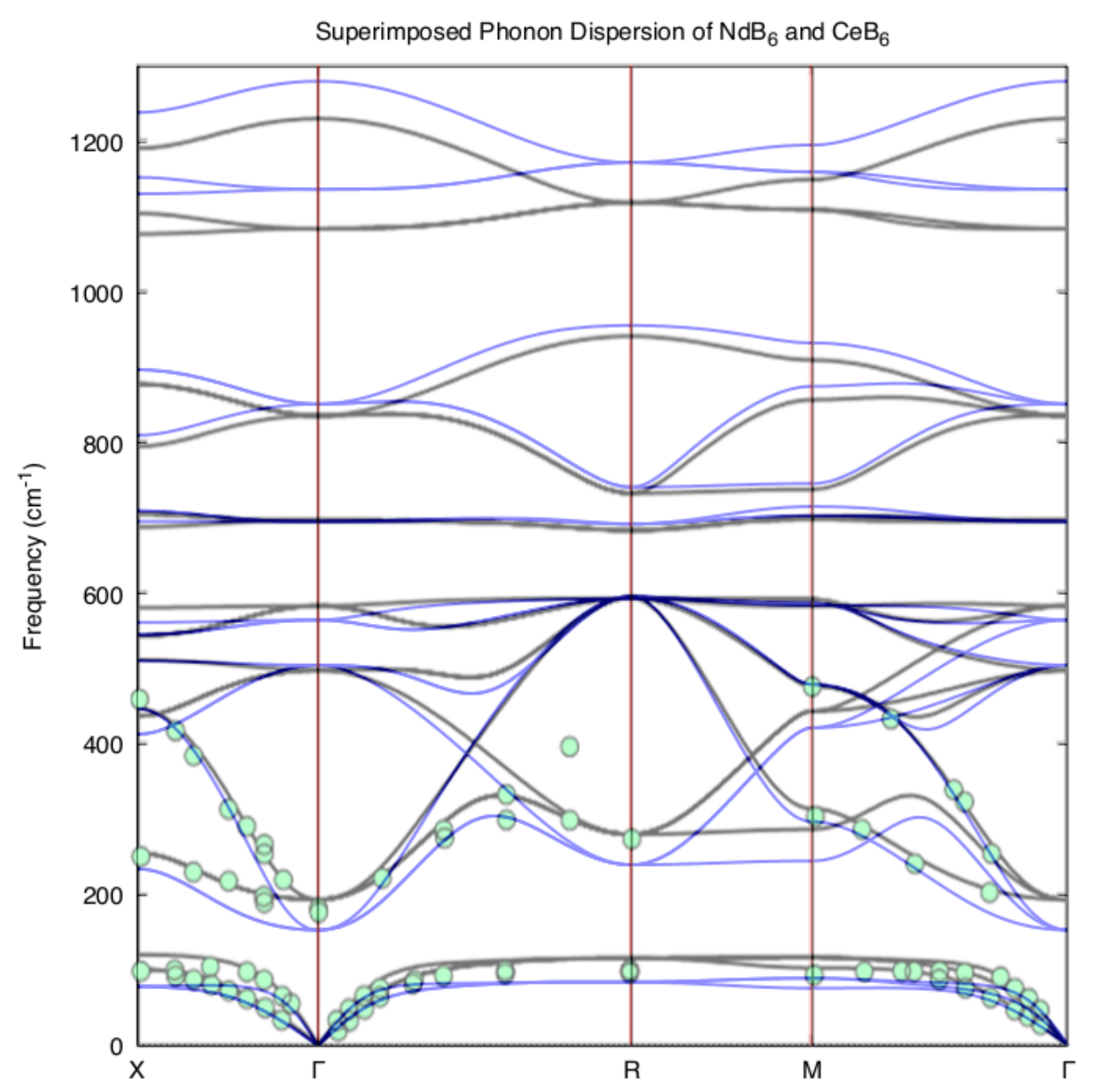}
    \caption{Comparison between calculated phonon dispersion of $1~\times~1~\times~1$ supercell NdB$_6$
      (blue) in this work and CeB$_6$. Two data sets for CeB$_6$ are shown,
      DFPT calculation results by G\"{u}rel and
      Eryi${\breve{\rm{g}}}$it~\cite{Gurel:2010} (black) and the
      experimental work of Kunii \etal{}~\cite{Kunii:1997}
      (circles). Modified from Figure 3 of G\"{u}rel and
      Eryi${\breve{\rm{g}}}$it's work~\cite{Gurel:2010}}
    \label{fig:CeB6}
  \end{center}
\end{figure}

\vspace{2mm}
The smaller primitive cell of NdB$_6$ allows for a larger, $2~\times~2~\times~2$ supercell to be
used in the phonon calculations. We would expect the best result for this supercell
relative to the other compounds within the scope of this work run
with less multiplied supercells.
In addition to the required $C_p$ data, we have also obtained the NdB$_6$ phonon
dispersion (Figure~\ref{fig:NdB6_phon}) in order to make a rough comparison with
other lanthanide series hexaborides.
Comparison is made between the phonon dispersion patterns of NdB$_6$ and LaB$_6$,
(Figure~\ref{fig:LaB6}), as well as NdB$_6$ and CeB$_6$ (Figure \ref{fig:CeB6}), both taken from
the work of G{\"{u}}rel and Eryi${\breve{\rm{g}}}$it.~\cite{Gurel:2010} It is evident
that the phonon dispersions are remarkably similar, which is expected due to the
similarity in their crystal structures, with exception for the lowest
and highest frequency phonon modes.
Nevertheless, all dispersion curves show a recognizable flat acoustic modes across the $q$-points,
which is indicative of freely vibrating large rare-earth ions in between octahedral B
atom "cages".
This feature is clearly seen in all the rare-earth hexaborides $R$B$_6$.

\begin{figure}
  \begin{center}
    \includegraphics[width=245pt]{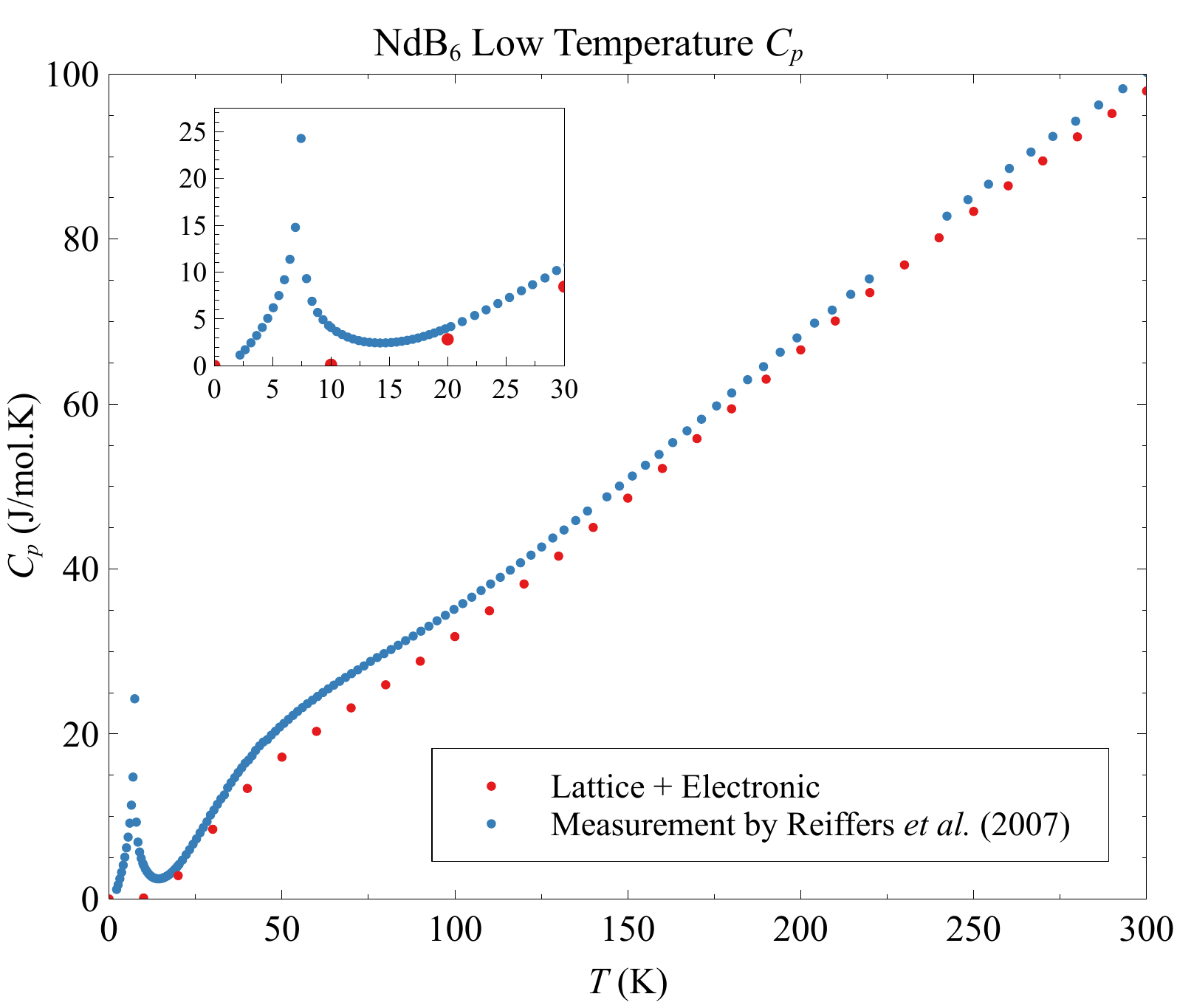}
    \caption{Comparison of calculated low temperature $C_p$ for NdB$_6$
      with measurement data from Reiffers \etal{}~\cite{Reiffers:2007}}
    \label{fig:NdB6lowCP}
  \end{center}
\end{figure}

\vspace{2mm}
The calculated $C_p$ of NdB$_6$ in this work
is in good agreement with the measurements of Reiffers~\etal{}, with the exception
in the very low temperature region $T < 20$ K (see Figure~\ref{fig:NdB6lowCP}).
Magnetic transitions that are not captured by QHA, largely contributes to $C_p$ in
this temperature region. Also seen in the low temperature region is a
Schottky anomaly-like bump starting around $T = 20$ K, which is also present
in other hexaborides such as YB$_6$~\cite{2006LOR} and CeB$_6$.~\cite{Gurel:2010}
This would later turn out to be significant in the application of our calculation
results to CALPHAD models, as discussed in Section~\ref{sec.CALPHAD}.

\begin{figure}
  \begin{center}
    \includegraphics[width=245pt]{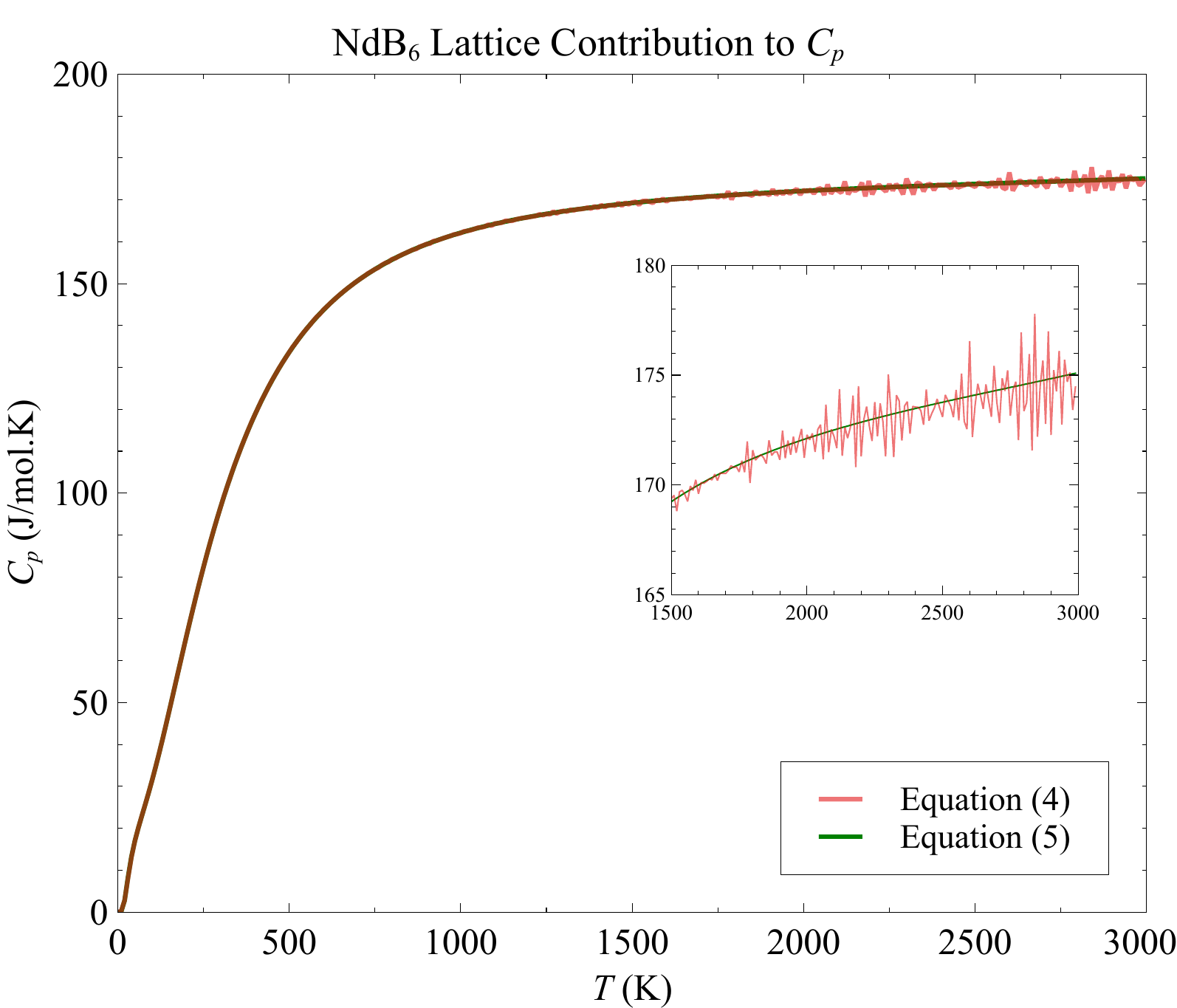}
    \caption{Comparison of two versions of calculated $C_p$ for NdB$_6$. Data points are replaced by line plot in order to emphasize fluctuation of results obtained from Equation~\eqref{Cp}}
    \label{fig:NdB6highCP}
  \end{center}
\end{figure}

\vspace{2mm}
We see in Figure~\ref{fig:NdB6highCP} that the formulation of $C_p$ from
Equation~\eqref{polyfit} is quantitatively similar to that of Equation~\eqref{Cp},
and is qualitatively less prone to numerical fluctuation. This leads us to believe
that Equation~\eqref{polyfit} is more suited to model $C_p$. Therefore, the $C_p$ derived from
Equation~\eqref{polyfit} is favored and is used from this point on in this work.

\begin{figure}
  \begin{center}
    \includegraphics[width=245pt]{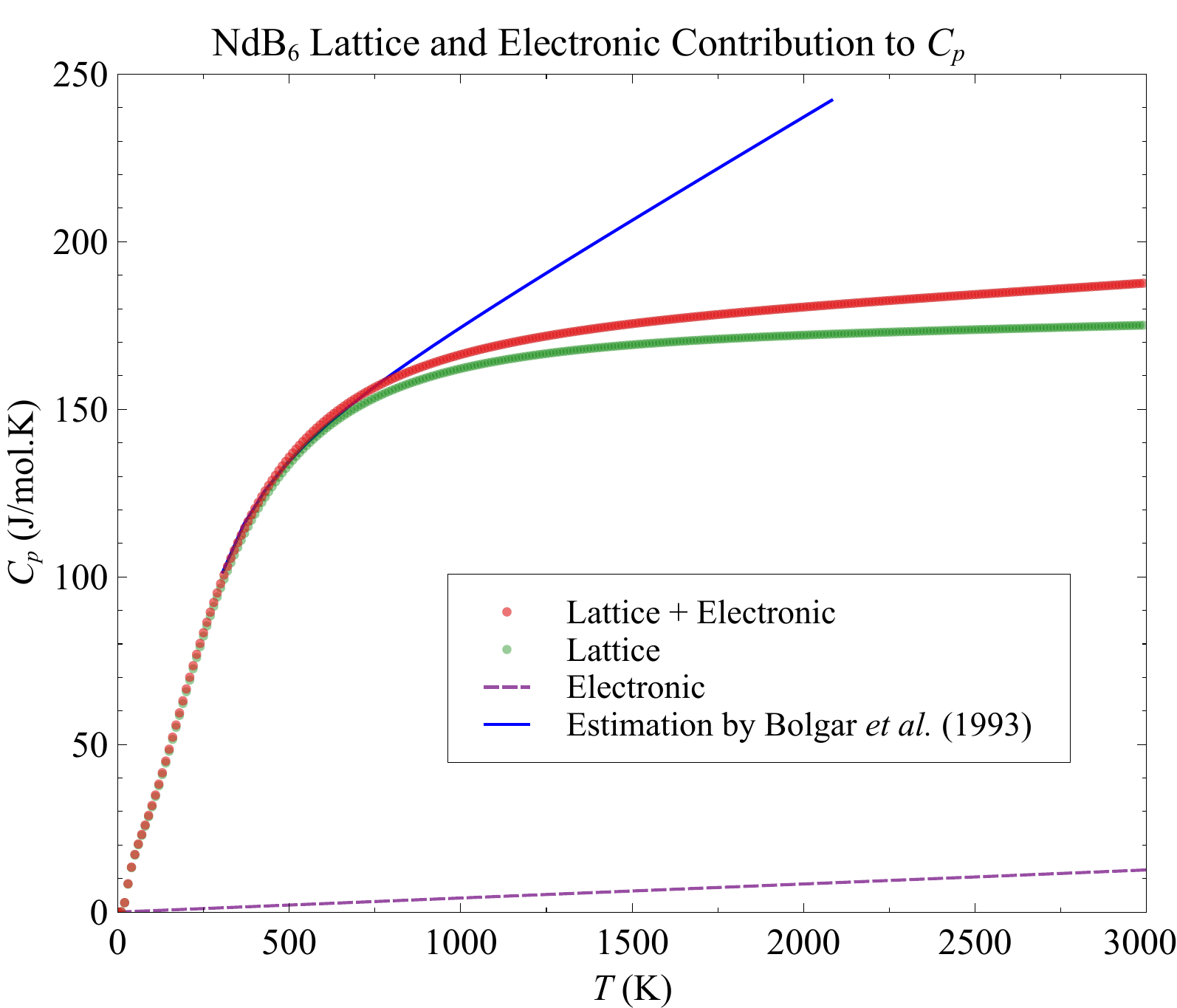}
    \caption{Contributions from both lattice vibrations and free electrons to the $C_p$ of NdB$_6$, with comparison with theoretical predictions data from Bolgar \etal{}~\cite{Bolgar:1993}}
    \label{fig:NdB6elec}
  \end{center}
\end{figure}

\vspace{2mm}
In order to pursue more accurate pictures for the heat capacity at higher temperatures,
we also performed calculations with much finer parameters from the previously obtained
ground state charge density in order to pursue a fine picture of the density of states
(DOS), specifically at the Fermi level.
From Equation~\eqref{elec}, the electronic contribution is added to the lattice contribution
in Equation~\eqref{polyfit}, resulting in a final $C_p$ data as shown in Figure~\ref{fig:NdB6elec}.
We made a comparison with theoretical prediction based on heat capacity
expression suggested by Bolgar \etal{}.~\cite{Bolgar:1993} Clearly, the calculated heat capacity diverges
cleanly from this theoretical estimation especially for the higher temperatures.
Though seemingly the divergence comes from a linear term for the $C_p$ figure,
adding an artificial linear correction only serves to greatly overestimate the low
$T$ region, suggesting that this disagreement is much more than a simple miscalculation
of the electronic contribution.
Therefore, our calculated values may provide significant contrast to earlier attempts
to assess the binary Nd-B phase diagram.~\cite{Hallemans:1995,Ende:2013,2019CHE}

\subsection{NdB$_4$}
\label{sec:NdB4}
The crystal structure of NdB$_4$ has the $P4/mbm$ symmetry. It
can regarded as a structure obtained from a distorted cubic NdB$_6$
lattice. The structure has B atom pairs between distorted cubic sublattices
of Nd atoms and B octahedras. Details concerning its crystal structure
are given in Table~\ref{tab:NdB4_str}. There is a single Nd Hubbard site in
the NdB$_4$ unit cell, which is the Nd 4$f$ orbital. Initial lattice parameters
and atomic positions are taken from the work of Salamakha~\etal{}.~\cite{Salamakha:2001}

\begin{table}
  \begin{center}
    \caption{Crystallographic data for NdB$_4$}
    \begin{tabular}{l l c c c c}
      \hline
      Space group && \multicolumn{4}{l}{$P4/mbm$ (127)}\\
	  Pearson symbol && \multicolumn{4}{l}{$tP20$}\\
      $a_0 [\AA]$ && \multicolumn{4}{l}{7.1775}\\
      $c_0 [\AA]$ && \multicolumn{4}{l}{4.0996}\\
      \hline
      \multirow{2}{*}{Label} & \multirow{2}{*}{Type} & \multicolumn{3}{c}{Fractional Coordinates} & \multirow{2}{*}{$U_{\rm{eff}}$ [eV]}\\
                  & & \phantom{0000}$x$\phantom{0000} & \phantom{0000}$y$\phantom{0000} & \phantom{0000}$z$\phantom{0000} &\\
      \hline
      Nd1 & Nd & 0.18306 & 0.68306 & 0.00000 & 5.7931\\
      B2  & B  & 0.00000 & 0.00000 & 0.20530 & -\\
      B3  & B  & 0.03924 & 0.17491 & 0.50000 & -\\
      B4  & B  & 0.58863 & 0.08863 & 0.50000 & -\\
      \hline
    \end{tabular}
    \label{tab:NdB4_str}
  \end{center}
\end{table}

\vspace{2mm}
The geometry optimization did not result in significant changes in the
lattice parameters and atomic positions from the initial values. The
calculated enthalpy of formation for NdB$_4$ are listed in
Table~\ref{tab:NdB4}, along with the experimental value from Meschel and
Kleppa.~\cite{Meschel:1995} In the GGA$+U+J$ calculations, the values of U
and J are taken from \cite{Nilsson:2013} as in the case of NdB$_6$. Similar to
NdB$_6$, GGA$+U+J$ and the simplified GGA$+U_{\rm{eff}}$ results are in
closer agreement with the experimental data. From these calculations it is
evident that simplified GGA$+U_{\rm{eff}}$ method is quite reliable.

\begin{table}
  \begin{center}
    \caption{Comparison of theoretical and experimental enthalpy of
      formation for NdB$_4$.}
    \begin{tabular}{l c}
      \hline
      Method & $\Delta _{\text{f}}H_{298}^\circ$ [J/mol] \\
      \hline
      GGA & -56482\phantom{ $\pm$ 1500}\\
      GGA$+U+J$\phantom{ $\pm$ 1.5} & -54937\phantom{ $\pm$ 1500}\\
      GGA$+U_{\rm{eff}}$ & -51613\phantom{ $\pm$ 1500}\\
      Experiment & -53300 $\pm$ 1500\\
      \hline
    \end{tabular}
    \label{tab:NdB4}
  \end{center}
\end{table}

\vspace{2mm}
Phonon calculations were performed using a $1 \times 1 \times 2$ supercell,
which is nearly cubic due to the $c/a$ ratio close to 0.56.
Unlike NdB$_6$, the use of a $2 \times 2 \times 2$ supercell would lead to
a very costly calculation process, not to mention the already sizable
cost for a frozen phonon calculation with a lower-symmetry structure.
The decision was made in light of this and taking into account the near
cubic properties of the $1 \times 1 \times 2$ supercell, which makes it
nearly ideal for a frozen phonon calculation (keeping effects from periodic
images at a near constant for each lattice vectors).

\begin{figure}
  \begin{center}
    \includegraphics[width=245pt]{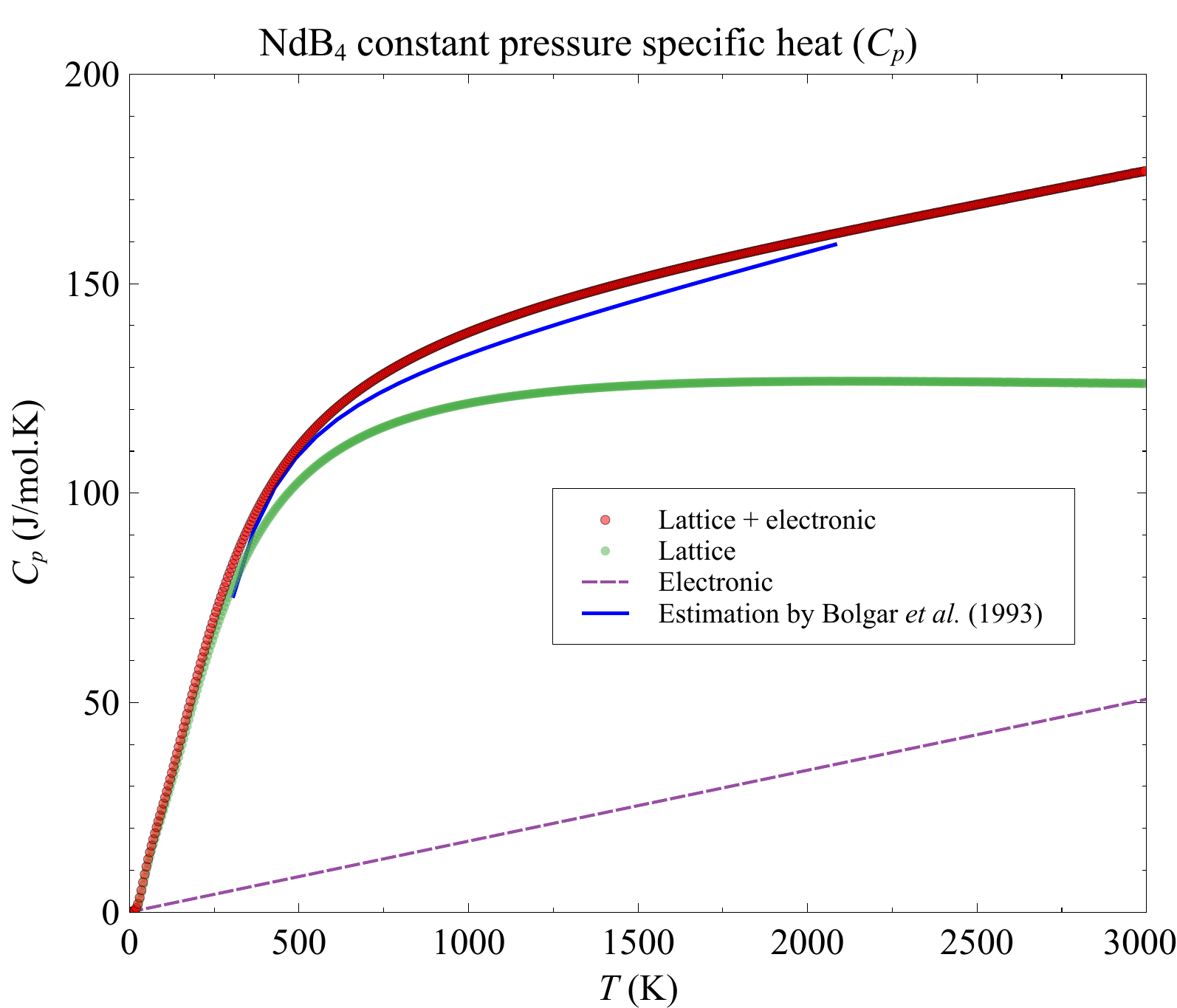}
    \caption{Calculated $C_p$ for NdB$_4$ and comparison with theoretical
      predictions from Bolgar~\etal{}.\cite{Bolgar:1993}}
    \label{fig:NdB4highCP}
  \end{center}
\end{figure}

\vspace{2mm}
Novikov \etal{}~\cite{Novikov:2011} studied the tetraborides of rare-earth
elements lanthanum, dysprosium, holmium, and lutetium for 2 K $< T <$ 300
K.  It was shown that the $C_p$ value for rare-earth tetraborides has a range of up to
80 J/mol/K at 300 K. The calculated $C_p$ for NdB$_4$ using QHA is shown in
Figure~\ref{fig:NdB4highCP}, and it corresponds to the
expected range for a rare-earth tetraboride as stated. These values are also
compared with the results of Bolgar \etal{}.~\cite{Bolgar:1993}

\vspace{2mm}
Another point of comparison may be taken from the low temperature measurement
performed by Watanuki \etal.~\cite{Watanuki:2009}
Similarly to the NdB$_6$ case, while fair agreement can be seen for $T > 20 K$,
small peaks in $T < 20$ K seen in the experimental results was not
reproduced in our calculations. As is with NdB$_6$, these peaks correspond to
the complex magnetic ordering in NdB$_4$ reported in other works, most
recently by Yamauchi \etal{}.~\cite{Yamauchi:2017}

\vspace{2mm}
Figure~\ref{fig:NdB4highCP} shows that our calculation result slightly overestimates
$C_{p}$ from the empirical estimations for $T > 300$ K.
However, it shows a very similar trend for higher temperatures with the Maier-Kelly
model from Bolgar~\etal{}~\cite{Bolgar:1993}, unlike the calculation results for NdB$_6$.
While showing the better agreement with theoretical prediction, the Schottky
anomaly-like bump for the $C_p$ figure also appears for low $T$, a similar
phenomena to NdB$_6$ and Nd$_2$B$_5$, which would later prove significant
in application to CALPHAD models.

\subsection{Nd$_2$B$_5$}
Crystal structure of Nd$_2$B$_5$ has a $C2/c$ symmetry with a
monoclinic unit cell, similar to Gd$_2$B$_5$. Initial lattice
parameters and atomic positions in our calculations are taken 
from the work of Roger \etal{}~\cite{Roger:2006}
(Table~\ref{tab:Nd2B5_str}). There are two Nd Hubbard sites in the
Nd$_2$B$_5$. The calculated enthalpy of formation of Nd$_2$B$_5$
is listed in Table~\ref{tab:Nd2B5}. Calculated enthalpies from GGA
and GGA$+U_{\rm{eff}}$ do not show good agreement with the
measurement by Meschel and Kleppa.~\cite{Meschel:2001}

\begin{table}
  \begin{center}
    \caption{Crystallographic data for Nd$_2$B$_5$.}
    \begin{tabular}{l l c c c c}
      \hline
      Space group && \multicolumn{4}{l}{$C2/c$ (127)}\\
      Pearson symbol && \multicolumn{4}{l}{$mP28$}\\
      $a_0 [\AA]$ && \multicolumn{4}{l}{15.0808}\\
      $b_0 [\AA]$ && \multicolumn{4}{l}{7.2522}\\
      $c_0 [\AA]$ && \multicolumn{4}{l}{7.2841}\\
      $\beta [^{\circ}]$ && \multicolumn{4}{l}{109.1040}\\
      \hline
      \multirow{2}{*}{Label} & \multirow{2}{*}{Type} & \multicolumn{3}{c}{Fractional Coordinates} & \multirow{2}{*}{$U_{\rm{eff}}$ [eV]}\\
                  & & \phantom{0000}$x$\phantom{0000} & \phantom{0000}$y$\phantom{0000} & \phantom{0000}$z$\phantom{0000} &\\
      \hline
      Nd1 & Nd & 0.11813 & 0.57013 & 0.59776 & 5.2099\\
      Nd2 & Nd & 0.11765 & 0.06276 & 0.72268 & 5.0492\\
      B3  & B  & 0.25110 & 0.78860 & 0.82890 & --\\
      B4  & B  & 0.25060 & 0.92270 & 0.04110 & --\\
      B5  & B  & 0.15980 & 0.75070 & 0.93930 & --\\
      B6  & B  & 0.75040 & 0.15990 & 0.40860 & --\\
      B7  & B  & 0.45800 & 0.25630 & 0.64720 & --\\
      \hline
    \end{tabular}
    \label{tab:Nd2B5_str}
  \end{center}
\end{table}

\begin{table}
  \begin{center}
    \caption{Comparison of theoretical and experimental enthalpy of
      formation for Nd$_2$B$_5$.}
    \begin{tabular}{l r}
      \hline
      Method & $\Delta _{\text{f}}H_{298}^\circ$ [J/mol] \\
      \hline
      GGA & -54933\phantom{ $\pm$ 1500}\\
      GGA$+U_{\rm{eff}}$ & -49938\phantom{ $\pm$ 1500}\\
      Experiment & -38900 $\pm$ 1500\\
      \hline
    \end{tabular}
    \label{tab:Nd2B5}
  \end{center}
\end{table}

\vspace{2mm}
Despite having only slightly distorted structure than NdB$_4$, the measured
enthalpy of formation of Nd$_2$B$_5$ are significantly different.
However, the computed values are not able to reflect this difference in enthalpy.
The similar results have been obtained in the recent DFT calculations for the Nd-B
binary system by Colinet and Tedenac~\cite{2018COL}, as well as for isostructural
Gd$_2$B$_5$ investigated in the same work. This discrepancy
between experimental and DFT results suggests a characteristic of the
$R_2$B$_5$ structure which is not being captured by either GGA or GGA$+U$
calculations.

\begin{figure}
  \begin{center}
    \includegraphics[width=245pt]{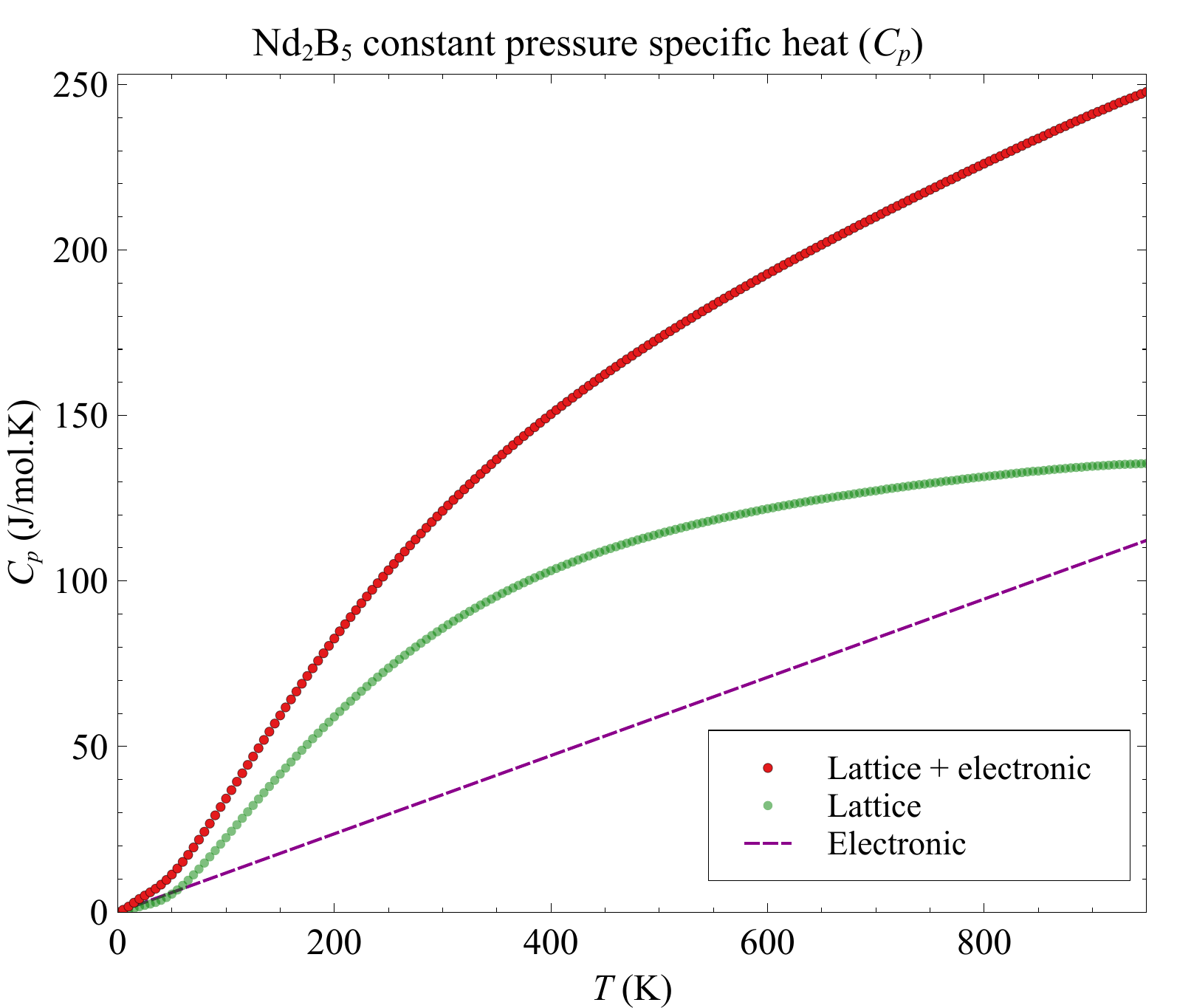}
    \caption{Calculated $C_p$ for Nd$_2$B$_5$ using QHA.}
    \label{fig:Nd2B5_cp}
  \end{center}
\end{figure}

\vspace{2mm}
The calculated $C_p$ for Nd$_2$B$_5$ is shown in Figure~\ref{fig:Nd2B5_cp}.
Phonon calculation was performed with only the primitive cell ($1 \times
 1 \times 1$) and as a result numerical instability is present for $T > 1000$ K.
Calculation results show the compound is very much metallic in nature, with
a large density of states at the Fermi level.
As a result, the electronic contribution is quite substantial and nearly equals the
lattice contribution for $T = 1000$ K.
However, as the enthalpy of formation has shown disagreement with experimental result,
it is perhaps wise to assign little to no weight for this data for CALPHAD purposes,
particularly due to the potential overbinding between lattices as a result of the
calculation.

\subsection{Nd$_2$Fe$_{17}$}
\label{sec:Nd2Fe17}
The unit cell of Nd$_2$Fe$_{17}$ belongs to the space group $R\bar{3}m$.
Initial lattice parameters were taken from Long \etal{}.~\cite{Long:1994}
Initial atomic positions are listed in the Pearson's Handbook of Crystallographic Data for Intermetallic Phases.~\cite{1985PEA}

\begin{table}
  \begin{center}
    \caption{Crystallographic data for Nd$_2$Fe$_{17}$.}
    \begin{tabular}{l l c c c c}
      \hline
      Space group && \multicolumn{4}{l}{$R\bar{3}m$ (166)}\\
      Pearson symbol && \multicolumn{4}{l}{$hR19$}\\
      $a_0 [\AA]$ && \multicolumn{4}{l}{8.5797}\\
      $c_0 [\AA]$ && \multicolumn{4}{l}{12.5021}\\
      \hline
      \multirow{2}{*}{Label} & \multirow{2}{*}{Type} & \multicolumn{3}{c}{Fractional Coordinates} & \multirow{2}{*}{$U_{\rm{eff}}$ [eV]}\\
                  & & \phantom{0000}$x$\phantom{0000} & \phantom{0000}$y$\phantom{0000} & \phantom{0000}$z$\phantom{0000} &\\
      \hline
      Nd1 & Nd & 0.00000 & 0.00000 & 0.33333 & 5.6000\\
      Fe2 & Fe & 0.00000 & 0.00000 & 0.09700 & 3.0130\\
      Fe3 & Fe & 0.33333 & 0.00000 & 0.00000 & 3.1539\\
      Fe4 & Fe & 0.50000 & 0.50000 & 0.16667 & 3.0244\\
      Fe5 & Fe & 0.50000 & 0.00000 & 0.50000 & 2.9752\\
      \hline
    \end{tabular}
    \label{tab:Nd2Fe17_str}
  \end{center}
\end{table}

\vspace{2mm}
Both Nd and Fe sites are treated as Hubbard sites, with five distinct
$U_{\rm{eff}}$ parameters for each site when considering all Nd and Fe atoms.
These parameters, as well as the crystal structure data, are listed in
Table~\ref{tab:Nd2Fe17_str}. This structure contains three formula units
(57 atoms). A rhombohedral primitive cell containing one unit formula (19 atoms)
is used for the calculations in this work. The magnetic moment in the primitive
cell is constrained to 51.5 $\mu _B$ according to the work of Long~\etal{}.~\cite{Long:1994}

\vspace{2mm}
The calculated enthalpy of formation for Nd$_2$Fe$_{17}$ and comparison
with experimental data is given in Table~\ref{tab:Nd2Fe17}. It is evident
that Hubbard correction results in significant deviations from regular GGA
calculations.  We apply the same treatment to the enthalpy of formation as
applied to the compound Nd$_5$Fe$_2$B$_6$ (see section~\ref{sec:Nd5Fe2B6}).

\begin{align}\label{eq:Eform1}
  \Delta _{\text{f}}H_{298}^\circ
  ({\text{Nd}_2}{\text{Fe}_{17}}) \approx\ &\ {E_{{\text{Nd}_2}{\text{Fe}_{17}},{\text{GGA+U(Nd)}}}}\nonumber\\
   & - 2{E_{{\text{Nd}},{\text{GGA+U}}}} - 17{E_{{\text{Fe}},{\text{GGA}}}}\\\nonumber
\end{align}


\vspace{2mm}
Unlike Nd$_5$Fe$_2$B$_6$, only treating Nd as Hubbard sites
(GGA+U$_{\rm eff}$, Nd) results in an overestimated value of enthalpy
of formation.  This implies that the bonds formed within the structure
is sufficiently delocalized such that the introduction of Hubbard $U$
correction does not properly reflect the electronic state. This suggests
that the $4f$ electrons in Nd participate in the bonding within
Nd$_2$Fe$_{17}$ sufficiently to lose its atomic-like nature.

\vspace{2mm}
The source of this behavior is attributed to the abundance of coordinated
Fe atoms around the Nd atoms.  In this compound, Nd atoms are present
in small amounts and always surrounded by 6 Fe atoms, a unique situation compared
to other compounds investigated in this study.  This might have led to a
hybridization between the $4f$ electron in Nd with one of the $3d$ levels in a
coordinated Fe atom.  Such a mechanism would explain how the loss of
atomic-like nature of the $4f$ orbitals and why the Hubbard U correction
does not reflect an accurate electronic structure.  This is a likely
explanation, especially given that Nd $4f$ electrons do slightly
contribute to bonding in metallic Nd~\cite{1971GSC}, in contrast with most
other lanthanides.

\vspace{2mm}
With this in mind, we calculated the enthalpy of formation of
Nd$_2$Fe$_{17}$ by regular GGA, and adding the effective Hubbard U
correction only for ground state Nd.  Instead of the uniform treatment of
Hubbard correction for all phases involved in the calculation
(GGA+U$_{\rm eff}$) or uniformly using regular GGA (GGA) results, only
ground state Nd is calculated with the Hubbard correction.

\begin{align}\label{eq:Eform2}
  \Delta _{\text{f}}H_{298}^\circ
  ({\text{Nd}_2}{\text{Fe}_{17}}) \approx\ &\ {E_{{\text{Nd}_2}{\text{Fe}_{17}},{\text{GGA}}}}\nonumber\\
   & - 2{E_{{\text{Nd}},{\text{GGA+U}}}} - 17{E_{{\text{Fe}},{\text{GGA}}}}\\\nonumber
\end{align}

\begin{figure}
  \begin{center}
    \includegraphics[width=245pt]{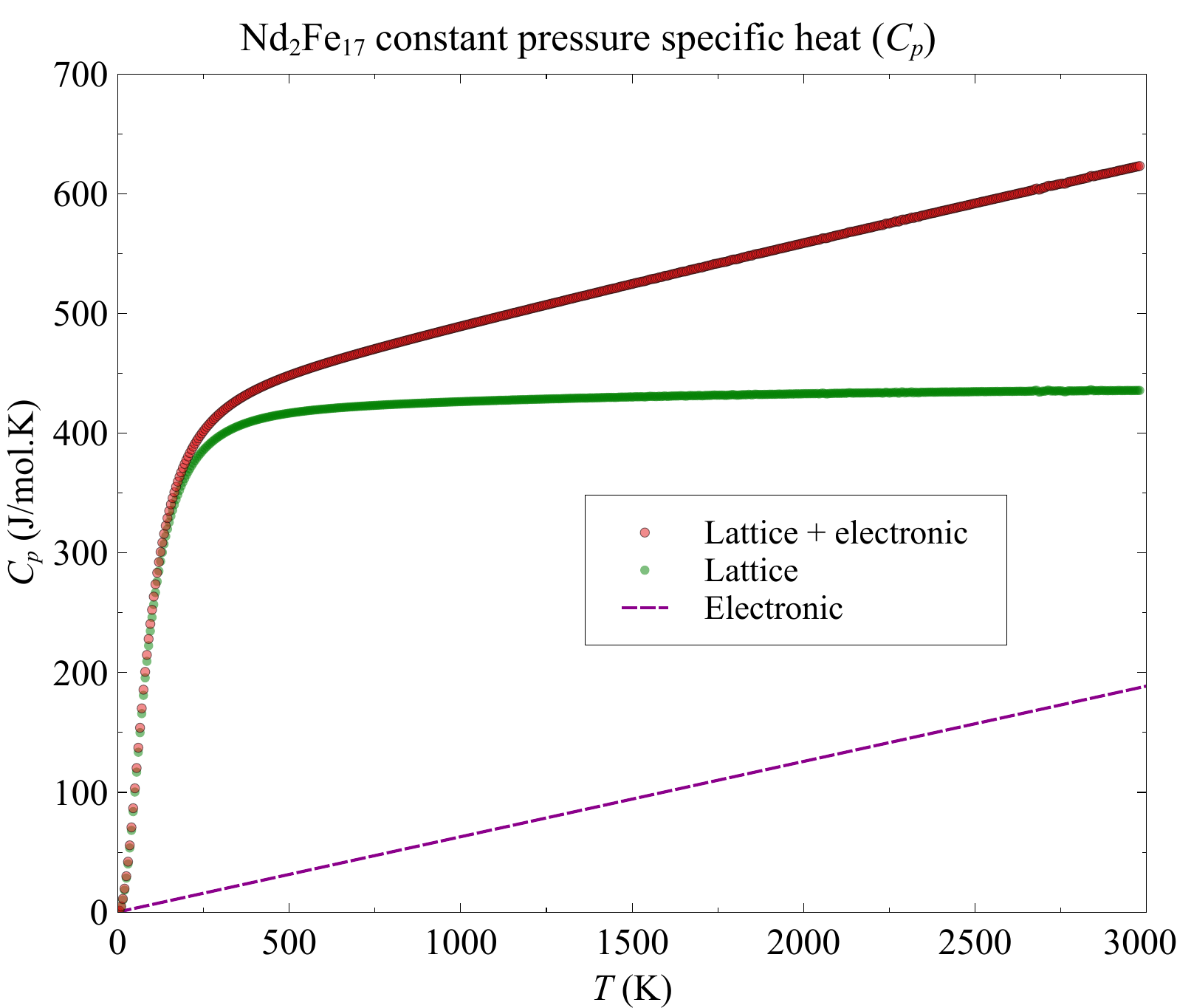}
    \caption{Calculated $C_p$ for Nd$_2$Fe$_{17}$ using QHA.}
    \label{fig:Nd2Fe17_cp}
  \end{center}
\end{figure}

\vspace{2mm}
This resulted in a slightly negative enthalpy of formation, as shown in
Table~\ref{tab:Nd2Fe17}.  It agrees well with the experimental data Meschel
\etal{}.~\cite{Meschel:2013} The calculated $C_p$ using QHA for this compound is shown
in Figure~\ref{fig:Nd2Fe17_cp}. As with Nd$_2$B$_5$, there are no experimental
measurements available for comparison.

\begin{table}
  \begin{center}
    \caption{Comparison of theoretical and experimental enthalpy of
      formation for Nd$_2$Fe$_{17}$.}
    \begin{tabular}{l r}
      \hline
      Method & $\Delta _{\text{f}}H_{298}^\circ$ [J/mol] \\
      \hline
      GGA & \phantom{$-1$}1410\phantom{ $\pm$ 15000}\\
      GGA$+U_{\rm{eff}}$ & -34428\phantom{ $\pm$ 15000}\\
      Equation~\eqref{eq:Eform1} & 18350\phantom{ $\pm$ 15000}\\
      Equation~\eqref{eq:Eform2} & \phantom{$1$}-3397\phantom{ $\pm$ 15000}\\
      Experiment & \phantom{0}-3000 $\pm$ 3900\phantom{0}\\
      \hline
    \end{tabular}
    \label{tab:Nd2Fe17}
  \end{center}
\end{table}

\subsection{Nd$_5$Fe$_2$B$_6$}
\label{sec:Nd5Fe2B6}
Nd$_5$Fe$_2$B$_6$ phase is one of three stable ternary compounds in Nd-Fe-B
system. It is also referred to as T3 phase.~\cite{Hallemans:1995} It has a
rhombohedral structure with $R\bar{3}m$ space group. A primitive cell with
13 atoms was used in the present work as input structure. The details about
the unit cell, containing 3 formula units (39 atoms), are summarized in
Table~\ref{tab:Nd5Fe2B6_str}. The initial lattice parameters and atom
positions were taken from Yartys~\etal{}.~\cite{Yartys:1997}

\begin{table}
  \begin{center}
    \caption{Crystallographic data for Nd$_5$Fe$_2$B$_6$.}
    \begin{tabular}{l l c c c c}
      \hline
      Space group && \multicolumn{4}{l}{$R\bar{3}m$ (166)}\\
      Pearson symbol && \multicolumn{4}{l}{$hR13$}\\
      $a_0 [\AA]$ && \multicolumn{4}{l}{5.4614}\\
      $c_0 [\AA]$ && \multicolumn{4}{l}{24.2720}\\
      \hline
      \multirow{2}{*}{Label} & \multirow{2}{*}{Type} & \multicolumn{3}{c}{Fractional Coordinates} & \multirow{2}{*}{$U_{\rm{eff}}$ [eV]}\\
                  & & \phantom{0000}$x$\phantom{0000} & \phantom{0000}$y$\phantom{0000} & \phantom{0000}$z$\phantom{0000} &\\
      \hline
      Nd1 & Nd & 0.00000 & 0.00000 & 0.25110 & 5.6556\\
      Nd2 & Nd & 0.00000 & 0.00000 & 0.41610 & 5.3697\\
      Nd3 & Nd & 0.00000 & 0.00000 & 0.00000 & 5.3833\\
      Fe4 & Fe & 0.00000 & 0.00000 & 0.12070 & 4.9680\\
      B5  & B  & 0.33333 & 0.00000 & 0.50000 & \phantom{0.0}-\phantom{00}\\
      \hline
    \end{tabular}
    \label{tab:Nd5Fe2B6_str}
  \end{center}
\end{table}

\begin{table}
  \begin{center}
    \caption{Nd$_5$Fe$_2$B$_6$ Enthalpy of Formation}
    \begin{tabular}{l c}
      \hline
      Method & $\Delta _{\text{f}}H_{298}^\circ$ [J/mol] \\
      \hline
      GGA & -40603\\
      GGA$+U_{\rm{eff}}$ & -77113\\
      Equation~\eqref{eq:Eform3} & -44686\\
      \hline
    \end{tabular}
    \label{tab:Nd5Fe2B6}
  \end{center}
\end{table}

\begin{figure}
  \begin{center}
    \includegraphics[width=245pt]{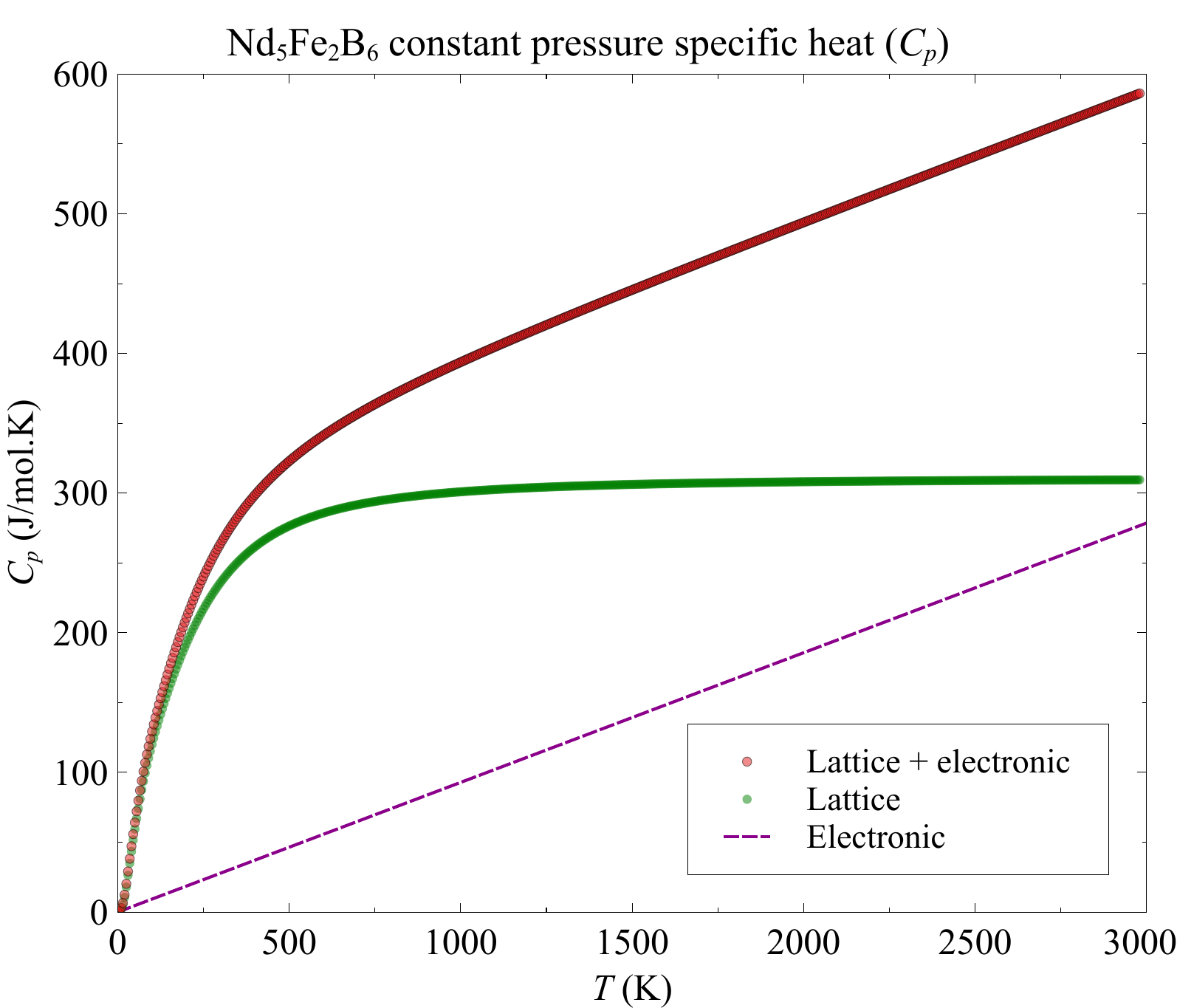}
    \caption{Calculated $C_p$ for Nd$_5$Fe$_2$B$_6$ using QHA.}
    \label{fig:Nd5Fe2B6_cp}
  \end{center}
\end{figure}

\vspace{2mm}
For this compound, implementing the Hubbard correction to Fe results in a
much larger shift of ground state energy for the elements and far less so
for the Nd$_5$Fe$_2$B$_6$ structure itself. This imbalance resulted in a
large shift in the calculated enthalpy of formation.  Another GGA+U
calculation on Nd$_5$Fe$_2$B$_6$ is performed by treating the Nd sites only
as Hubbard sites.  We take the value of U$_{\rm eff}$ for Fe to be zero.

\begin{align}\label{eq:Eform3}
  \Delta _{\text{f}}H_{298}^\circ
  ({\text{Nd}_5}{\text{Fe}_2}{\text{B}_6}) \approx\ &\ {E_{{\text{Nd}_5}{\text{Fe}_2}{\text{B}_6},{\text{GGA+U(Nd)}}}}- 5{E_{{\text{Nd}},{\text{GGA+U}}}}\nonumber\\
   & - 2{E_{{\text{Fe}},{\text{GGA}}}} - 6{E_{{\text{B}},{\text{GGA}}}}\\\nonumber
\end{align}


\vspace{2mm}
The calculated enthalpy of formation is listed in
Table~\ref{tab:Nd5Fe2B6}. It can be seen that the modified
GGA$+U_{\text{eff}}$ gives better agreement with the enthalpy obtained from
GGA-PBE value. The inclusion of Fe as Hubbard sites overestimates the
enthalpy value. The calculated $C_p$ for Nd$_5$Fe$_2$B$_6$ using QHA is
shown in Figure~\ref{fig:Nd5Fe2B6_cp}. However, without any available
experimental results for comparison, caution must be exercised in its
eventual use in the CALPHAD modelling of the system.

\section{CALPHAD}
\label{sec.CALPHAD}
\begin{figure*}
  \begin{center}
    \includegraphics[width=300pt]{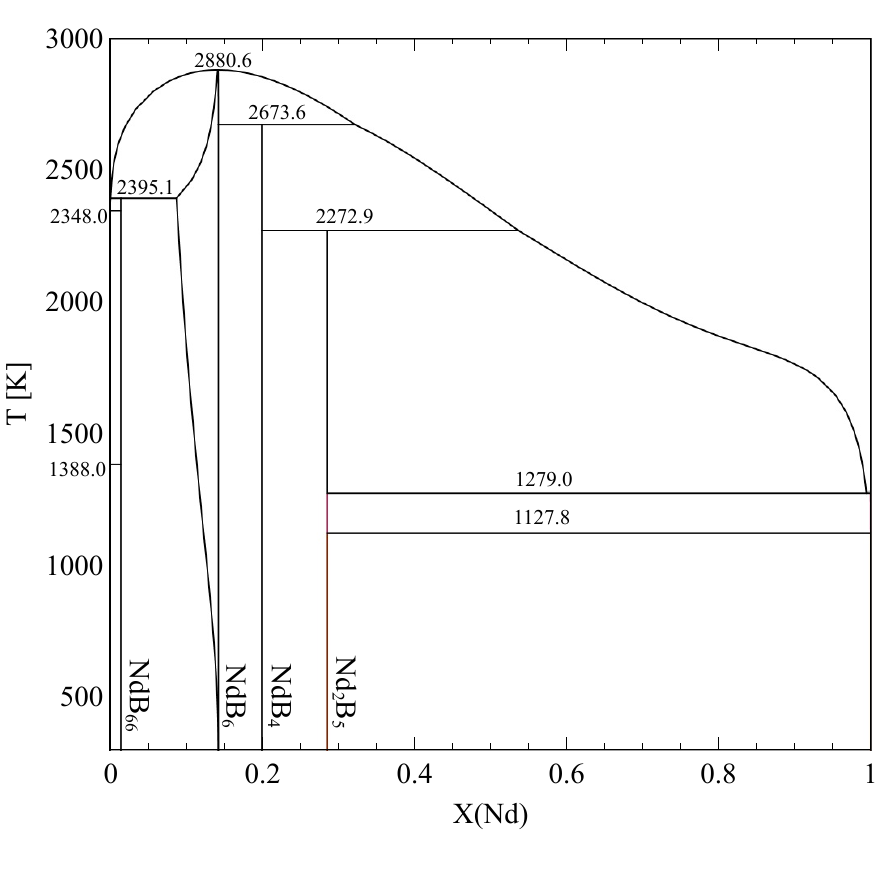}
    \caption{Calculated phase diagram of the Nd-B binary system for $T > 300$ K, utilizing {\emph{ab initio}} calculation results in this work}
    \label{fig:PD}
  \end{center}
\end{figure*}

{\emph{Ab initio}} calculation results obtained were used as data for the
optimization of the binary Nd-B phase diagram.
As a good starting point, we utilized the results of the assessment of
Chen~\etal{}~\cite{2019CHE} which solely utilized available experimental data.
The Gibbs energy parameters present in this previous work were further optimized
using the calculation results from the Nd-B compounds investigated in this work.
In this way, the our optimization takes into account all the available experimental
data so far and supplements them with our {\emph ab initio} calculation.
The results of our successful re-optimization of the Gibbs energy model parameters
for Nd-B binary system is shown in Figure~\ref{fig:PD}.

\begin{figure}
  \begin{center}
    \includegraphics[width=245pt]{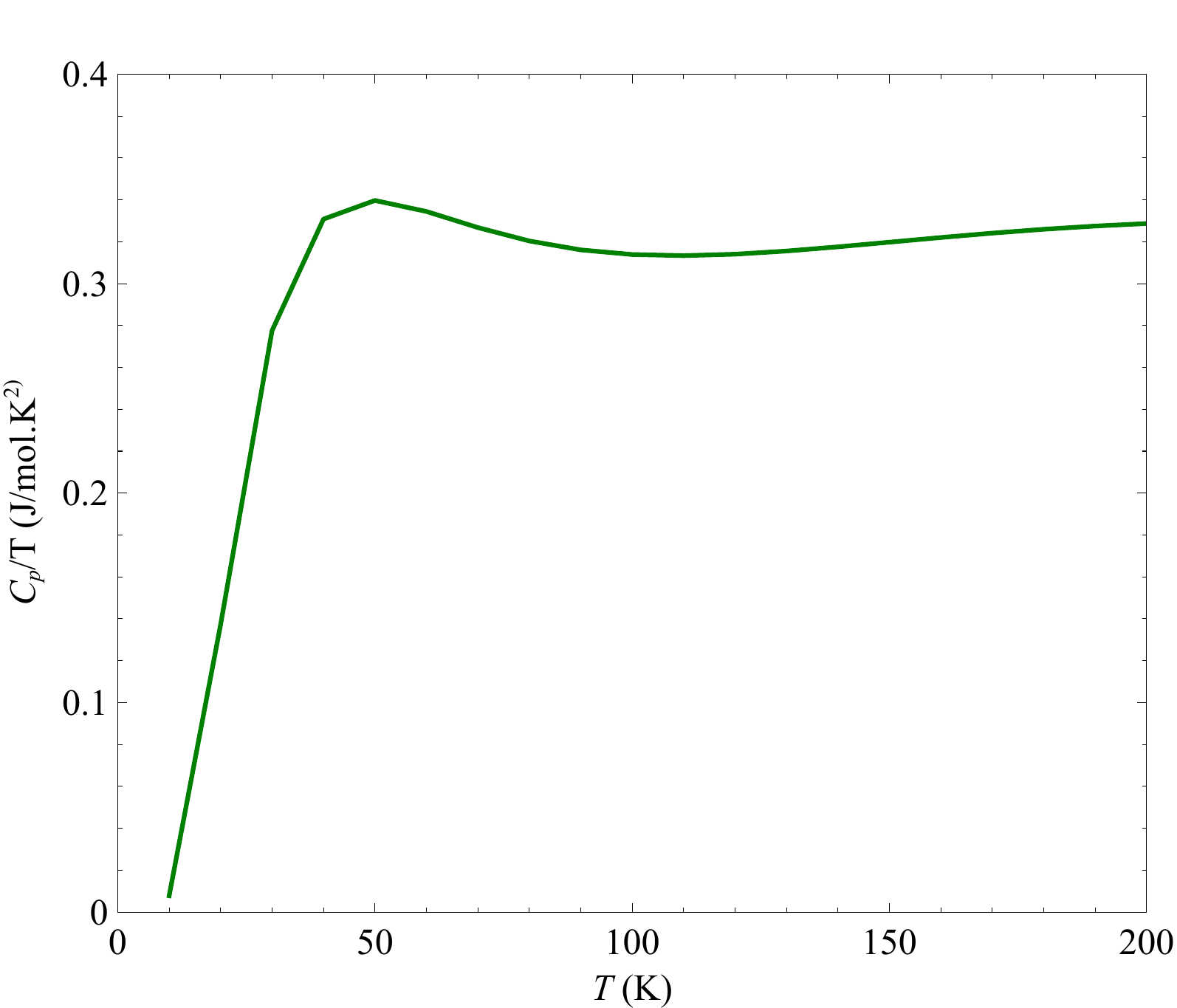}
    \caption{Calculated $C_p/T$ plot for NdB$_6$. An unusual peak is seen around $T = 50$ K, followed
	by the more expected linear section.}
    \label{fig:anomaly}
  \end{center}
\end{figure}

\vspace{2mm}
Several issues emerged during our re-optimization, most notably due to the Schottky-like
anomaly present in the $C_p$ data for binary Nd-B compounds, seen in Figures~\ref{fig:NdB6elec},
\ref{fig:NdB4highCP}, \ref{fig:Nd2B5_cp}, and in more detail in the $C_p/T$ plot in
Figure~\ref{fig:anomaly} for NdB$_6$.
As the Schottky anomaly is often traced back to changes in entropy due to spin populations, which is
unaccounted for in phonon calculations (vibrational/lattice contribution to $C_p$), it is unlikely
that these broad peaks are due to magnetic effects at higher $T$.
Instead, it has been attributed to a large dependence of lattice volume on low temperature
phonon modes as in the case of YB$_6$~\cite{2006LOR}, and it is likely that the same anomaly
encountered in our calculations is due to the same property in NdB$_6$.

\vspace{2mm}
Nevertheless, the existence of this anomaly in the $C_p$ data proved to inhibit the fitting of a
$C_p$ function for NdB$_6$ and NdB$_4$ for the full range of $0 < T < 3000$ K.
This, and the fact that Gibbs energy functions for unary compounds (Nd and B) are unavailable for
$T < 300$ K, led us to optimize the phase diagram only in the $T > 300$ K range, as seen in
Figure~\ref{fig:PD}.
$C_p$ data of NdB$_6$ was not fitted to a function; instead, the discrete data points are combined
with all available experimental data points, and subsequently used to optimize directly the
entropic term for the Gibbs energy models for NdB$_6$.
The calculated NdB$_4$ heat capacity turned out to inhibit instead of improving the optimization process,
and as such was discarded.
This is attributed to the insufficient parameters in the calculation to reliably obtain good $C_p$ data,
which proves to be less robust in the optimization than expected.
Our results do not differ much from that of our initial starting point for optimization from
Chen~\etal{}~\cite{2019CHE}, with the significant exception for the area of the plot around
NdB$_6$, to which we have applied the enthalpy of formation and $C_p$ data from our calculation.

\section{Summary}
\label{sec.conc}
The summary of this work can be found in the following Table~\ref{tab:summary}.

\begin{table}[h]
  \begin{center}
    \caption{Summary of enthalpy of formation}
    \begin{tabular}{l c c}
      \hline
      \multirow{2}{*}{Compound} & $\Delta _{\text{f}}H_{298}^\circ$ [J/mol] & \multirow{2}{*}{Literature [J/mol]}\\
      & (this work) &\\
      \hline
      NdB$_6$ & -45019 & -46750 $\pm$ 1250~\cite{Storms:1981}\\
      NdB$_4$ & -51613 & -53300 $\pm$ 1500~\cite{Meschel:1995}\\
      Nd$_2$B$_5$ & -49938 & -38900 $\pm$ 1500~\cite{Meschel:2001}\\
      Nd$_2$Fe$_{17}$ & \phantom{$1$}-3397 & \phantom{0}-3000 $\pm$ 3900~\cite{Meschel:2013}\\
      Nd$_5$Fe$_2$B$_6$ & -44686 & -\phantom{00}\\
      \hline
    \end{tabular}
    \label{tab:summary}
  \end{center}
\end{table}

\vspace{2mm}
We have performed DFT calculations on several stable binary and ternary
compounds in Nd-Fe-B and its constituent binary systems in order to obtain
values of enthalpy of formation and $C_p$. The simplified Hubbard $U$
correction $U_{\rm eff}$ was adopted in order to account for the localized
$4f$ orbitals present. The results of our calculations are compared with
theoretical predictions and experimental measurements from literature. It
is evident that Hubbard $U$ correction, treating Nd atoms as Hubbard sites,
successfully corrects for the regular GGA method on the prediction of
enthalpy of formation for several compounds.

\vspace{2mm}
The choice of Hubbard sites is another point we can draw from our work.
We observe the inclusion of Fe as Hubbard sites results in large deviations from
the actual measured value in experiments performed (Tables~\ref{tab:Nd2Fe17}
and~\ref{tab:Nd5Fe2B6}). In both cases massive overbinding can be observed
in the calculations, suggesting that GGA+U severely underestimates the energy
of bulk Fe, which subsequently greatly affects Fe-rich compounds.
Treating Nd as a Hubbard site is as well not always accurate, at least for
compounds with exceedingly low concentration of Nd, as coordination with
neighbouring atoms may affect the atomic-like nature of 4$f$ electrons in Nd.
In such cases, treating Nd with regular GGA is sufficient as seen in the case
for Nd$_2$Fe$_{17}$ in this work.

\vspace{2mm}
Phonon calculations were performed in order to obtain lattice vibrations
contribution to $C_p$. It was seen that the results achieve good agreement with
low temperature measurements except for magnetic ordering in $T < 20$ K.
The $C_p$ data obtained for Nd-B binary compounds show a Schottky-like
anomaly in the low $T$ region of around $T = 50$ K, which is due to the sudden
spike in the phonon density of states in the same temperature energy region.
This anomaly interfered with the fitting of calculated $C_p$ data to the
usual functions regularly used in other materials, and as such the integration
of this data required a different approach laid out in Section~\ref{sec.CALPHAD}.

\vspace{2mm}
Our calculated results for enthalpies of formation and $C_p$ data from NdB$_6$
and the enthalpy of formation for NdB$_4$ are subsequently used for a re-optimization
of Gibbs energy models from a previous work by Chen~\etal{}.~\cite{2019CHE}
This re-optimized phase diagram for the binary Nd-B system was successfully obtained
for $T > 300$ K with improved results specifically for compounds NdB$_6$ and NdB$_4$,
as expected, due to the integrated calculation results from this work.

\section{Acknowledgements}
Authors are grateful to Prof. Marian Reiffers of the Faculty of Humanities
and Natural Sciences, Presov University, Presov, Slovakia, for fulfilling
our request for the measurement results of the heat capacity of NdB$_6$ in
one of his previous works.
The computation in this work has been performed using the facilities of
the Research Center for Advanced Computing Infrastructure (RCACI) at JAIST.
A.T.H. is grateful for financial supports from MEXT-KAKENHI
(JP16K21724). R.M. is grateful for financial supports from MEXT-KAKENHI
(17H05478 and 16KK0097), from FLAGSHIP2020 (project nos. hp180206 and
hp180175 at K-computer), from Toyota Motor Corporation, from I-O DATA
Foundation, and from the Air Force Office of Scientific Research
(AFOSR-AOARD/FA2386-17-1-4049).  K.H. is grateful for financial supports
from a Grant-in-Aid for Scientific Research on Innovative Areas (16H06439
and 19H05169), PRESTO (JPMJPR16NA) and the ``Materials
research by Information Integration Initiative" (MI$^2$I) project of the
Support Program for Starting Up Innovation Hub from Japan Science and
Technology Agency (JST).

\section{Data availability}
Both the raw and processed data required to reproduce these findings are available to download from http://dx.doi.org/10.17632/n4f563wf7m.2

\bibliography{references}

\appendix
\setcounter{secnumdepth}{0}
\section{APPENDIX: Computational Details}
  \label{append}
  The following constitutes the details for the computational calculations
  performed within this work. As outlined in Section~\ref{sec.det.spec},
  convergence of the ground state total energy is utilized in order to
  determine optimal calculation parameters.  We observe a convergence of
  ground state total energy to 0.001~Ry/atom to be sufficient to establish
  sufficiently large cut-off energy and sufficiently dense $k$-mesh.  As
  for the width of the Marzari-Vanderbilt smearing, a looser criteria of
  0.002~Ry/atom is adopted in order to facilitate calculation convergence
  more easily.

  \subsection{Nd}
  \noindent
  Optimized calculation parameters are as follows:\\[2mm]
  \begin{tabular}{l l}
    Cut-off energy (wavefunction) & : 40 Ry\\
    Cut-off energy (electron density) & : 320 Ry\\
    Maximum smearing width & : 0.04 Ry\\
    $k$-mesh & : $10\times 10\times 3$\\[2mm]
  \end{tabular}

  \noindent
  The ground state of Nd converges to antiferromagnetic ordering.
  This is also confirmed by \emph{ab initio} calculation on our part,
  by comparing the ground state energy result of the antiferromagnetic
  ordering and ferromagnetic ordering; the antiferromagnetic structure
  results in a lower ground state energy.
  GGA+$U_{\rm eff}$ geometry optimization of the ground state crystal structure
  produced:

  \begin{table}[h]
   \begin{center}
    \caption{Geometry optimization results for ground state $\alpha$-Nd}
     \begin{tabular}{l r r}
      \hline
       & Optimized & Experiment\\
      \hline
      $a [\AA]$ & 3.7532 & 3.6582\\
      $c [\AA]$ & 12.6511 & 11.7966\\
      $\mu$ [$\mu _B$/cell] & 0.06 & 0.00\\
      \hline
     \end{tabular}
    \label{tab:apNd}
   \end{center}
  \end{table}

  \subsection{Fe}
  \noindent
  Optimized calculation parameters are as follows:\\[2mm]
  \begin{tabular}{l l}
    Cut-off energy (wavefunction) & : 40 Ry\\
    Cut-off energy (electron density) & : 320 Ry\\
    Maximum smearing width & : 0.01 Ry\\
    $k$-mesh & : $4\times 4\times 4$\\[2mm]
  \end{tabular}

  \noindent
  The ground state of Fe converges to ferromagnetic ordering. As
  previously stated, GGA+$U_{\rm eff}$ geometry optimization of
  ground state Fe results in an overestimation of the lattice constant
  to 2.9127~\AA. Meanwhile, GGA geometry optimization results in a
  slight underestimation of the lattice constant of 2.8253~\AA, in
  good agreement with previous \emph{ab initio} works which utilized
  the GGA-PBE exchange correlation functional.~\cite{2011JAN,2010KOD}

  \begin{table}[h]
   \begin{center}
    \caption{Geometry optimization results for ground state $\alpha$-Fe}
     \begin{tabular}{l r r}
      \hline
       & Optimized & Experiment\\
      \hline
      $a [\AA]$ & 2.8253 & 2.8665\\
      $\mu$ [$\mu _B$/cell] & 2.08 & 2.20\\ 
      \hline
     \end{tabular}
    \label{tab:apFe}
   \end{center}
  \end{table}

  \subsection{B}
  \noindent
  Optimized calculation parameters are as follows:\\[2mm]
  \begin{tabular}{l l}
    Cut-off energy (wavefunction) & : 35 Ry\\
    Cut-off energy (electron density) & : 280 Ry\\
    $k$-mesh & : $4\times 4\times 4$\\[2mm]
  \end{tabular}

  \noindent
  B atoms are not counted as Hubbard sites and are non-magnetic.
  As such, geometry optimization are performed as a non-spin polarized
  calculation with regular GGA, with the results shown in
  Table~\ref{tab:apB}.

  \begin{table}[h]
   \begin{center}
    \caption{Geometry optimization results for ground state $\alpha$-B}
     \begin{tabular}{l r r}
      \hline
       & Optimized & Experiment\\
      \hline
      $a [\AA]$ & 4.8978 & 4.9179\\
      $c [\AA]$ & 12.5412 & 12.5805\\
      \hline
     \end{tabular}
    \label{tab:apB}
   \end{center}
  \end{table}

  \subsection{NdB$_6$}
  \noindent
  Optimized calculation parameters are as follows:\\[2mm]
  \begin{tabular}{l l}
    Cutoff energy (wavefunction) & : 40 Ry\\
    Cutoff energy (electron density) & : 320 Ry\\
    $k$-mesh & : $9\times 9\times 9$\\
    Maximum smearing width & : 0.01 Ry\\[2mm]
  \end{tabular}

  \noindent
  Unlike ground state Nd, we found for NdB$_6$ ferromagnetic ordering
  to be energetically more favorable compared to antiferromagnetic ordering.
  GGA+$U_{\rm eff}$ geometry optimization produces results shown in
  Table~\ref{tab:apNdB6}

  \begin{table}[h]
   \begin{center}
    \caption{Geometry optimization results for NdB$_6$}
     \begin{tabular}{l r r}
      \hline
       & Optimized & Experiment\\
      \hline
      $a [\AA]$ & 4.1391 & 4.126\\
      $\mu$ [$\mu _B$/cell] & 0.00 & 0.00\\
      \hline
     \end{tabular}
    \label{tab:apNdB6}
   \end{center}
  \end{table}

  \subsection{NdB$_4$}
  \noindent
  Optimized calculation parameters are as follows:\\[2mm]
  \begin{tabular}{l l}
    Cut-off energy (wavefunction) & : 40 Ry\\
    Cut-off energy (electron density) & : 380 Ry\\
    Maximum smearing width & : 0.01 Ry\\
    $k$-mesh & : $5\times 5\times 9$\\[2mm]
  \end{tabular}

  \noindent
  GGA+$U_{\rm eff}$ geometry optimization results are shown
  in Table~\ref{tab:apNdB4}.
  
  \begin{table}[h]
   \begin{center}
    \caption{Geometry optimization results for NdB$_4$}
     \begin{tabular}{l r r}
      \hline
       & Optimized & Experiment\\
      \hline
      $a [\AA]$ & 7.2384 & 7.1775\\
      $c [\AA]$ & 4.1212 & 4.0996\\
      $\mu$ [$\mu _B$/cell] & 0.00 & 0.00\\
      \hline
     \end{tabular}
    \label{tab:apNdB4}
   \end{center}
  \end{table}

  \subsection{Nd$_2$B$_5$}
  \noindent
  Optimized calculation parameters are as follows:\\[2mm]
  \begin{tabular}{l l}
    Cut-off energy (wavefunction) & : 40 Ry\\
    Cut-off energy (electron density) & : 480 Ry\\
	Maximum smearing width & : 0.02 Ry\\
    $k$-mesh & : $4\times 4\times 5$\\[2mm]
  \end{tabular}

  \noindent
  GGA+$U_{\rm eff}$ geometry optimization results are shown
  in Table~\ref{tab:apNd2B5}.
  
  \begin{table}[h]
   \begin{center}
    \caption{Geometry optimization results for Nd$_2$B$_5$}
     \begin{tabular}{l r r}
      \hline
       & Optimized & Experiment\\
      \hline
      $a [\AA]$ & 15.1832 & 15.0808\\
      $b [\AA]$ & 7.2384 & 7.2522\\
      $c [\AA]$ & 7.2696 & 7.2841\\
      $\beta [^{\circ}]$ & 109.567 & 109.1040\\
      $\mu$ [$\mu _B$/cell] & 25.39 & \multicolumn{1}{c}{-}\\
      \hline
     \end{tabular}
    \label{tab:apNd2B5}
   \end{center}
  \end{table}

  \subsection{Nd$_2$Fe$_{17}$}
  \noindent
  Optimized calculation parameters are as follows:\\[2mm]
  \begin{tabular}{l l}
    Cut-off energy (wavefunction) & : 125 Ry\\
    Cut-off energy (electron density) & : 1200 Ry\\
    Maximum smearing width & : 0.03 Ry\\
    $k$-mesh & : $4\times 4\times 4$\\[2mm]
  \end{tabular}

  GGA geometry optimization results are shown in
  Table~\ref{tab:apNd5Fe2B6}. Nd$_2$Fe$_{17}$ magnetization
  is constrained to 51.5 $\mu _B$/cell.~\cite{Long:1994}
  
  \begin{table}[h]
   \begin{center}
    \caption{Geometry optimization results for Nd$_2$Fe$_{17}$}
     \begin{tabular}{l r r}
      \hline
       & Optimized & Experiment\\
      \hline
      $a [\AA]$ & 8.3927 & 8.5797\\
      $c [\AA]$ & 12.7008 & 12.5021\\
      $\mu$ [$\mu _B$/cell] & 51.5 & 51.5\\
      \hline
     \end{tabular}
    \label{tab:apNd2Fe17}
   \end{center}
  \end{table}

  \subsection{Nd$_5$Fe$_2$B$_6$}
  \noindent
  Optimized calculation parameters are as follows:\\[2mm]
  \begin{tabular}{l l}
    Cut-off energy (wavefunction) & : 85 Ry\\
    Cut-off energy (electron density) & : 850 Ry\\
    Maximum smearing width & : 0.03 Ry\\
    $k$-mesh & : $3\times 3\times 3$\\[2mm]
  \end{tabular}

  GGA+$U_{\rm eff}$ geometry optimization results are shown in
  Table~\ref{tab:apNd5Fe2B6}.
  
  \begin{table}[h]
   \begin{center}
    \caption{Geometry optimization results for Nd$_5$Fe$_2$B$_6$}
     \begin{tabular}{l r r}
      \hline
       & Optimized & Experiment\\
      \hline
      $a [\AA]$ & 5.5559 & 5.4614\\
      $c [\AA]$ & 25.3520 & 24.2720\\
      $\mu$ [$\mu _B$/cell] & 1.13 & \multicolumn{1}{c}{-}\\
      \hline
     \end{tabular}
    \label{tab:apNd5Fe2B6}
   \end{center}
  \end{table}

\end{document}